\documentclass[]{article}
\usepackage{bm}
\usepackage{xcolor}
\usepackage{cancel}
\usepackage{amsmath}
\usepackage{amsthm}
\usepackage{bbm}
\usepackage{mathtools}
\usepackage{nicematrix}
\usepackage{amssymb}
\usepackage{apacite}
\usepackage{natbib}
\usepackage{graphicx}
\usepackage{caption}
\usepackage{multirow}
\usepackage{float}
\usepackage{array}
\usepackage[symbol]{footmisc}
\usepackage[caption = false]{subfig}
\usepackage[a4paper, total={6.5in, 9in}]{geometry}
\usepackage{adjustbox}
\usepackage{lscape}
\usepackage[colorlinks=true,citecolor=blue,allcolors=blue]{hyperref} 

\title{\textbf{A Bayesian actor-oriented multilevel relational event model with hypothesis testing procedures}}

\author{}

\author{{Fabio Vieira}$^{1}$ \and Roger Leenders$^{2,3}$ \and Daniel McFarland$^4$ \and Joris Mulder$^1$}
\date{%
    $^1$Department of Methodology and Statistics, Tilburg University\\%
    $^2$Department of Organization Studies, Tilburg University \\
    $^3$Jheronimus Academy of Data Science\\
    $^4$Department of Education, Stanford University\\[2ex]%
}

\newtheorem{lemma}{Lemma}
\theoremstyle{plain}
\newtheorem*{theorem*}{Theorem}

\newcolumntype{P}[1]{>{\centering\arraybackslash}p{#1}}

\NiceMatrixOptions{
code-for-first-row = \color{black} ,
code-for-last-row = \color{black} ,
code-for-first-col = \color{black} ,
code-for-last-col = \color{black}
}

\setcitestyle{aysep={}} 
\providecommand{\keywords}[1]
{
  \small	
  \textbf{Keywords:} #1
}

\begin{document}

\maketitle

\begin{abstract}
Relational event network data are becoming increasingly available. Consequently, statistical models for such data have also surfaced. These models mainly focus on the analysis of single networks, while in many applications, multiple independent event sequences are observed, which are likely to display similar social interaction dynamics. Furthermore, statistical methods for testing hypotheses about social interaction behavior are underdeveloped. Therefore, the contribution of the current paper is twofold. First, we present a multilevel extension of the dynamic actor-oriented model, which allows researchers to model sender and receiver processes separately. The multilevel formulation enables principled probabilistic borrowing of information across networks to accurately estimate drivers of social dynamics. Second, a flexible methodology is proposed to test hypotheses about common and heterogeneous social interaction drivers across relational event sequences. Social interaction data between children and teachers in classrooms are used to showcase the methodology.
\end{abstract}


\section{Introduction}

\par
Current technological advancements, combined with constant development of new communication applications, has originated huge amounts of data on social interactions \citep{eagle2003social}. 
This has enabled researchers to build and test increasingly rich and complex 
models of human interaction using time-stamped data. Social network analysis is among the fields that have benefited the most from having access to rich temporal interaction data. In this paper, we specifically focus on the Relational Event network model \citep{butts2017relational, stadtfeld2017interactions, mulder2019modeling}

\par
A relational event network is comprised of multiple interactions among a finite set of actors. Observation units are called ``relational events", which are defined as discrete instances of interactions among social entities along a timescale \citep{butts2017relational}. In a \textit{directed} network, events contain a clear indication of sender and receiver. Thus, every observation unit displays information on which actor was the sender, which actor was the receiver, and at which point in time the interaction occurred. The stream of events in a given network is often referred to as a relational event sequence.

\par
One of the traditional approaches to social network analysis is based on Markovian random graphs theory and involves the aggregation of events into a graph \citep{van2009random}. Transitions on the graph structure are then modeled via sufficient statistics, which are functions of the observed network \citep*{frank1986markov, hanneke2010discrete, lusher2013exponential}. A different approach was taken by \cite{butts2008}, who employed survival analysis concepts and introduced the \textit{relational event model} which profits from the temporal structure of relational events. In this framework, the main goal is to model the rates of communication between sender and receivers via a log-linear function, without requiring the data to be aggregated. This is done by employing sufficient statistics that capture important social patterns and actor-specific covariates. 

\par
Since then, the relational event model has gained increasing popularity and received multiple extensions. For instance, \cite{vu2011continuous} introduced a model with time-varying parameters, based on the additive Aalen model \citep{aalen1989linear}, and developed an algorithm for online inference in social networks. 
\cite{perry2013point} used a partial likelihood approach to modeling who the receiver will be for a given sender. \cite{vu2015relational} implemented case-control sampling to decrease the number of computations in estimating relational event models. Later, the control-case sampling approach was further explored by \cite{lerner2020reliability} to estimate relational event models in large networks. \cite*{stadtfeld2017dynamic} and \cite{stadtfeld2017interactions} introduced a two-step model, where the first step consists of modeling the activity rate of the sender and the second one features the choice of the receiver conditional on the sender. \cite{mulder2019modeling} proposed a way to analyze the temporal evolution of effects in the social network, by estimating effects in different subsets of the data determined by overlapping intervals, which they called moving windows, this approach was further developed by Meijerink et al. (2022).
\par
These contributions have mainly focused on the statistical analysis of a single relational event sequence (with a notable exception of \cite{dubois2013hierarchical}) while in practice we also often observe multiple independent relational event sequences which show similar but not identical social interaction behavior. For example, \cite{blonder2011time} analyses information flow in multiple relational event networks of ant colonies, \cite{dubois2013hierarchical} studies several relational event sequences of high-school students interactions and \cite{kiti2016quantifying} examines social contact in numerous networks of Kenyan households. The gold standard for analyzing such clustered (or hierarchical) data is using a multilevel or mixed effects approach. In a relational event modeling approach, this implies that a probability distribution is specified for the network effects (or drivers) of social interaction behavior (such as inertia, reciprocity, transitivity, or group effects) across clusters. A multilevel approach will result in `pooled' estimates of the cluster-specific network effects where information about social interaction behavior in other clusters is borrowed to improve the estimation \citep{gelman2006data}.
We achieve this by modeling the relational event sequences in a Bayesian hierarchical framework via Markov chain Monte Carlo (MCMC) methods. 

\par
Therefore, our first contribution is a multilevel extension for the dynamic actor-oriented model \citep{stadtfeld2017dynamic} for independent relational event sequences. In this framework, a relational event between a sender and a receiver in a network is modeled by separately modeling (i) when an actor decides to initiate an event as a sender given the past history, and (ii) modeling the receiver that is chosen given the sender and the past event history between the actors. Thereby, this approach differs from the dyadic approach of \cite{butts2008}, and the hierarchical extension by \cite{dubois2013hierarchical}, where the time, sender, and receiver are jointly modeled using the same set of parameters \citep{stadtfeld2017rejoinder}. The actor-oriented approach is therefore more flexible by separately modeling the behavior of actors to become a sender and the behavior of senders to choose a receiver in contrast to the dyadic approach. A second important difference is that we explicitly model the full unstructured covariance matrix of the network effects across sequences. Thereby the dependency between interaction behavior across sequences, such as inertia and reciprocity (which is generally nonzero), is included in the analysis resulting in improved estimation using the concept of pooling. A final important difference we note here between a dyadic relational event model and an actor-oriented approach is the computational burden, which is considerably larger in the case of a dyadic approach. This can be explained by the size of the risk set (i.e., the possible events that can be observed at every step), which is equal to $N(N-1)$ in the case of a network consisting of $N$ actors where $N(N-1)$ dyads are at risk for each event, while the risk set of the sender model in an actor-oriented model consists of $N$ actors, and the risk set of the receiver model consists of $N-1$ actors (when assuming that the sender cannot be equal to the receiver). In the case of moderate to large networks, this difference can have a huge impact on the computational costs when using a dyadic approach.


\par
Our second contribution is an extensive set of statistical tests using the Bayes factor \citep{jeffreys1961theory,kass1995bayes} for evaluating hypotheses about social interaction behavior in the case of independent relational event sequences. The first test can be used to assess whether specific interaction behavior, such as inertia (which quantifies the degree of habitual behavior between actors to keep sending messages of one actor towards another actor), is equal or differs across sequences (i.e., whether it should be modeled as a fixed effect or as a random effect). In the case of different interaction behavior across sequences, a second test is proposed to assess the degree of heterogeneity (or variability) of network effects on social interaction behavior, e.g., to assess whether inertia behavior varies more across clusters than reciprocity, or whether the impact of a group effect (e.g., whether the sender is the teacher in a class room or not) on social interaction dynamics or more or less heterogeneous than another group effect (e.g., whether the sender is male or female). Third, a test is proposed for evaluating hypotheses with equality and order constraints on the average relative importance across clusters (i.e., on the fixed effects) of drivers on social interaction dynamics. The methodology is flexible to also allow for testing the absolute values of network effects, which is useful when the exact sign is unknown before observing the data. These tests build on and further extends recent developments on Bayes factor hypothesis testing \citep{mulder2019bfpack}.

\par
Our methods have been implemented in R \citep{R2017}, interfacing with Stan through the rstan package \citep{standev2018rstan}. We provide the full code to facilitate the fitting of other relational event models (see \textcolor{red}{link suppressed for peer review}).

\par
The remainder of this paper is divided as follows: in Section \ref{sec:model} we describe the relational events framework and present the hierarchical actor-oriented relational event model. This Section also introduces the Bayesian specification of the model by presenting prior distributions and briefly discussing a reparametrization that allows more efficient sampling; we detail the hypothesis testing methods in Section \ref{sec:hyp_test} and conduct the empirical application in Section \ref{sec:applic}. The paper ends with a discussion in Section \ref{sec:disc}, where we go through some limitations of the model and possible routes for future research.

\section{The relational event framework}\label{sec:model}

The relational event model (REM) is used to model the rate of interactions in dynamic social networks \citep{butts2008}. In this framework, events happen among actors at particular points in time, being represented by tuples in the form $e_t = (s, r, t)$, where $s$ is the sender, $r$ is the receiver, and $t$ is the time point of the interaction. Thus $e_t$ is called a relational event. Therefore, we have $K$ clusters (in this paper we will use cluster, groups, relational event sequence and network interchangeably), each of which with $N_1, \dots, N_K$ social actors allowed to be senders or receivers at any given time. At time $t$, in cluster $k$, sender $s \in \{1, 2, \dots, N_{k} \}$ interacts with receiver $r \in \{1, 2, \dots, N_{k}\ |\ r \ne s \}$, forming a dyad $(s, r) \in \mathcal{R}_{k}(t)$, where $\mathcal{R}_{k}(t) = \big\{(s , r): s, r \in \{1, 2, \dots, N_{k}\},\ s \ne r \big\}$  is called the risk set, which comprises all possible dyads $(s, r)$ at a particular point in time.

\par
The main idea consists of assuming that, for each cluster $k,\ \text{for}\ k = 1, \dots, K$, we are able to observe an ordered sequence of $M_{k}$ dyadic events among $N_{k}$ individuals on the time window $[0, \tau_{k}) \in {\rm I\!R}^+$. Thus a relational event sequence for cluster $k$ is formally defined as

\begin{equation*}
    \textbf{E}_{k} = \{e_{t_{m}} = (s_{m}, r_{m}, t_{m}):\ (s_{m}, r_{m}) \in \mathcal{R}_{k}(t_{m}),\ 0 < t_1 < \dots < t_{M_k} < \tau_{k}\},
\end{equation*}

\noindent
where $t_m$ is the time at which the $m^{th}$ event occurred, in cluster $k$. Then, following \cite{butts2008}, who borrows concepts from survival analysis, the relational event history $\textbf{E}_{k}$ is modeled as a stochastic process, where the rate of events from sender $s$ to receiver $r$ , with $(s, r) \in \mathcal{R}_{k}(t)$, is given by the intensity $\lambda_{s r} \big(t | \textbf{E}_{k} \big)$. This intensity function has the form of a Cox proportional hazards model \citep{cox1972regression}. Moreover, the intensity is assumed constant between subsequent events and the waiting times conditionally exponentially distributed. This amounts to the well know piecewise constant exponential model for survival data \citep{friedman1982piecewise}. Thus, the survival function is given by $S_{s r} \big(t_{m} - t_{m-1} | \textbf{E}_{k} \big) = \exp \big\{ - (t_{m} - t_{m-1}) \lambda_{s r} \big(t_{m} | \textbf{E}_{k} \big) \big\}$.

\subsection{The actor-oriented multilevel relational event model}\label{AREM}

\par
In this paper, we focus on an alternative relational event model that was proposed by \cite{stadtfeld2017dynamic} and \cite{stadtfeld2017interactions}. Their approach models the receiver given the sender, similar to \cite{perry2013point}. Conceptually, it builds on the same tradition as the stochastic actor-oriented model \citep{snijders1996stochastic}, where the evolution of the network is assumed to be a product of actors' individual behaviors as they constantly seek to maximize their own utilities. The basic framework consists of a two-step approach based on log-linear predictors. First, the waiting time until an actor becomes active is modeled. After that, a multinomial choice model \citep{mcfadden1973conditional} is employed to determine the choice of the receiver by the active sender actor. These two steps are assumed to be conditionally independent given the available information about the past up to that event.

\par
The Bayesian hierarchical approach we will develop for multiple relational event sequences (denoted as "clusters") will build on this model. At each point in time, in cluster $k$, sender $s \in \{1, 2, \dots, N_{k} \}$ starts an interaction with intensity $\lambda_{s} \big(t | \textbf{E}_{k} \big)$. This intensity is directly proportional to the probability of a given actor $s$ to be the next sender. This probability is given by $P\big(s_m = s | \textbf{E}_{k}\big) = \lambda_s\big(t | \textbf{E}_{k}\big)/\sum_{h \in k} \lambda_h\big(t | \textbf{E}_{k}\big),\ \forall\ h \in \{1, 2, \dots, N_{k} \}\ \text{and}\ m \in \{1, 2, \dots, M_{k}\}$, where $M_{k}$ is the number of events in cluster $k$. Next, this sender chooses the receiver $r \in \{1, 2, \dots, N_{k} | r \neq s \}$, forming the dyad $(s, r) \in \mathcal{R}_{k}(t)$, with intensity $\lambda_{r | s} \big(t | s, \textbf{E}_{k} \big)$. The receiver intensity represents the rate at which actor $s$ chooses actor $r$ to form a dyad, which is proportional to the probability of observing dyad $(s, r) \in \mathcal{R}_{k}(t)$ as the next one in the sequence. This probability is given by $P\Big(r_m = r | s_m = s, \textbf{E}_{k}\Big) = \lambda_{r|s}\Big(t | s, \textbf{E}_{k}\Big)/\sum_{h \in k} \lambda_{h|s}\Big(t | s, \textbf{E}_{k}\Big),\ \forall\ h \in \{1, 2, \dots, N_{k}\ |\ h \neq s \}\ \text{and}\ m \in \{1, 2, \dots, M_{k}\}$. Then, these intensities for cluster $k$, for the sender and receiver steps of this model, will be given by the following log-linear functions:

\begin{equation}
    \begin{split}
    \lambda_{s}\big(t | \textbf{E}_{k}\big) = &\exp \{ \bm{\phi}' \bm{z}_{s}(t) + \bm{\gamma}'_k \bm{x}_{s}(t)\},\\
    \lambda_{r|s} \big(t | s, \textbf{E}_{k} \big) = &\exp \{\bm{\psi}' \bm{z}_{s r}(t) + \bm{\beta}_{k}' \bm{x}_{s r}(t)\},
    \end{split}
\end{equation}

\noindent
where $\bm{\phi}$ and $\bm{\psi}$ are a vectors of fixed-effect parameters, $\bm{\gamma}_k$ and $\bm{\beta}_k$ are a vectors of random-effect parameters for cluster $k$, with $k = 1, \dots, K$. The vectors of statistics $\bm{z}_{s}(t)$ and $\bm{x}_{s}(t)$ are associated with actor $s \in \{1, 2, \dots, N_{k} \}$, whereas $\bm{z}_{sr}(t)$ and $\bm{x}_{sr}(t)$ are vectors of statistics associated with dyad $(s, r) \in \mathcal{R}_{k}(t)$.

\par
Assuming this log-linear form will allow us to conduct inference at the actor level, unveiling effects that make actors more (or less) prone to start interactions or more (or less) likely to be chosen as the next receiver. Also, this idea helps us limit the size of the set of possible dyads that need to be analyzed at each point in time: if actor $s$ is the one starting an interaction at time $t$, then all dyads where $s$ is not the sender become impossible to happen. Thus, for the actor-oriented model, the risk set, $\mathcal{R}_{k}(t)$, will have size $N_{k}-1$ (given the sender), whereas for a dyadic model, such as in \cite{dubois2013hierarchical}, the risk set has size $(N_{k}-1)N_{k}$, which can easily become massive for medium to large networks.

\par
The likelihood for the actor-oriented model is given by

\begin{equation} \label{actor_like}
    \begin{split}
      p(\textbf{E} | \bm{\theta}, \bm{Z}, \bm{X}) = \prod_{k = 1}^{K} &\prod_{m = 1}^{M_{k}} \Bigg[ \Bigg( \lambda_{s_m}\Big(t_m | \textbf{E}_{k}\Big) \prod_{s \in k} S_s\Big(t_m - t_{m-1} | \textbf{E}_{k}\Big) \Bigg) \times \\   &\times \frac{\lambda_{r_m | s_m}\Big(t_m| s_m, \textbf{E}_{k}\Big)}{\sum_{r \in k} \lambda_{r | s_m}\Big(t_m| s_m, \textbf{E}_{k}\Big)}  \Bigg]  \times \prod_{s \in k} S_{s}\Big(\tau_{k} - t_{M_{k}} | \textbf{E}_{k}\Big),
    \end{split}
\end{equation}

\noindent
where $\bm{\theta}$ is the vector containing all parameters and $\bm{Z}$ and $\bm{X}$ are matrices with fixed and random effects covariates, respectively. The time of the last observed event in cluster $k$ is denoted by $t_{M_{k}}$ and $\tau_{k}$ the end of the observation period. In most empirical applications, it is assumed that $t_{M_{k}} = \tau_{k}$. This way the last part of the likelihood is equal to 1, since due to the piecewise constant exponential assumption we have $S_{s_{m}} \big(t_{m}  - t_{m-1} | \textbf{E}_{k} \big) = \exp \big\{ - (t_{m} - t_{m-1}) \lambda_{s} \big(t_{m} | \textbf{E}_{k} \big) \big\}$. In this setting, at time $t_m$, $P\big(s = s_m | \textbf{E}_{k}\big) = \frac{\lambda_s}{\sum_{h \in k} \lambda_h},\ \forall\ s \in k$ is the probability of sender $s$ being active. The probability of actor $r$ being the receiver given that $s$ is the sender is given by $P\big(r_m = r | s_m = s, \textbf{E}_{k}\big) = \frac{\lambda_{r|s}}{\sum_{h \in k} \lambda_{h|s}},\ \forall\ r \in k$.

\par
The likelihood in equation \eqref{actor_like} is a product of the two pieces. The first one, representing the sender model, is a piece-wise constant exponential likelihood

\begin{equation*}
    p_{\text{sender}}(\textbf{E} | \bm{\theta}, \bm{Z}, \bm{X}) = \prod_{k = 1}^{K} \prod_{m = 1}^{M_{k}} \Bigg[ \Bigg( \lambda_{s_m}\Big(t_m | \textbf{E}_{k}\Big) \prod_{s \in k} S_s\Big(t_m - t_{m-1} | \textbf{E}_{k}\Big) \Bigg) \Bigg]
\end{equation*}

\noindent
and the second one, representing the receiver model, is a multinomial likelihood

\begin{equation*}
    p_{\text{receiver}}(\textbf{E} | \bm{\theta}, \bm{Z}, \bm{X}) = \prod_{k = 1}^{K} \prod_{m = 1}^{M_{k}} \Bigg[ \frac{\lambda_{r_m | s_m}\Big(t_m| s_m, \textbf{E}_{k}\Big)}{\sum_{r \in k} \lambda_{r | s_m}\Big(t_m| s_m, \textbf{E}_{k}\Big)} \Bigg].
\end{equation*}

\noindent
In the literature, it has been shown that both of these models are special cases of the Poisson model. \cite{holford1980analysis} and \cite{laird1981covariance} are examples of cases where the Poisson representation of the piece-wise exponential model is discussed. \cite{baker1994multinomial} extensively discusses the multinomial-Poisson transformation and how their likelihoods yield identical estimates. For proofs, see Appendix \ref{app:AO-Poisson}. The Poisson regression model has been extensively studied, and it is well understood in the statistical literature \citep{frome1983analysis, consul1992generalized, hayat2014understanding}.

The second level of the hierarchical actor-oriented relational event model specifies the multivariate distributions of network effects across the $K$ sequences. We follow the standard approach in multilevel modeling by assuming multivariate normal distributions for $\bm\beta_k$ under the sender model and for $\bm\gamma_k$ under the receiver model, i.e.,
\begin{eqnarray*}
\bm{\beta}_k &\sim& \mathcal{N} (\bm{\mu}_{\beta}, \bm{\Sigma}_{\beta})\\
\bm{\gamma}_k &\sim& \mathcal{N}(\bm{\zeta}_{\gamma}, \bm{\Sigma}_{\gamma}),
\end{eqnarray*}
for $k = 1, \dots, K$. The mean vectors quantify the overall, global effect of the network effects and the unstructured covariance matrices quantify the variability of the effects across clusters and the dependency structure between the effects across clusters. Thus, when estimating these distributions when fitting the multilevel model using independent relational event sequences, the estimated means and (co)variances of the second level are used to improve the estimates of the specific parameters in the separate sequences (especially in the case of short sequences).

\subsection{Prior specification}\label{sec:bayes_inf}
In a Bayesian approach, prior distributions have to be specified which reflect our uncertainty about the model parameters before observing the data. Throughout this paper we shall work with vague, noninformative priors (which are completely dominated by the data).

\par
The prior choice of the random effects covariance matrices is most important. 
We decompose the random effects covariance matrix under the receiver model according to 
\begin{equation}\label{cov_decomp}
    \bm{\Sigma}_{\beta} = \bm{\sigma}_{\beta} \times \bm{\Omega}_{\beta} \times \bm{\sigma}_{\beta},
\end{equation}

\noindent
where $\bm{\sigma}_{\beta} \coloneqq diag(\sigma_{\beta,1}, \dots, \sigma_{\beta,P})$ is a diagonal matrix of standard deviations and $\bm{\Omega}_{\beta}$ is a correlation matrix \citep{gelman2006data}. Following \cite{carpenter2017stan}, $\bm{\Omega}$ will have a Lewandowski-Kurowicka-Joe (LKJ) prior and $\bm{\tau}$ a half-Cauchy prior as follows
\begin{itemize}
    \item $\sigma_{\beta,p} \sim \text{half-Cauchy}(0, \tau_{\beta}),\ p = 1,\dots, P,$
    \item $\bm{\Omega}_{\beta} \sim \text{LKJCorr}(\eta_{\beta}).$
\end{itemize}
It has been shown that a half-Cauchy prior for the random effects standard deviation results in desirable estimates in the case of multilevel data with few clusters \citep{gelman2006data}. Furthermore, by setting very large prior scale parameters, we can obtain approximately flat priors for the standard deviations. Further note that a half-Cauchy prior for the standard deviation corresponds to a $F$ distribution on the variance \citep{mulder2018matrix}, which is a common distribution for modeling variance components.

\noindent
The LKJ prior is defined as $\text{LKJCorr}(\bm{\Omega} | \eta) \propto \text{det}(\bm{\Omega})^{\eta - 1}$, with $\eta \in {\rm I\!R}^{+}$. This distribution allows us to sample uniformly from the space of positive definite correlation matrices and has a behavior similar to the beta distribution \citep*{wang2018equivalence, lewandowski2009generating}. For example, when $\eta = 1$ it has a uniform behavior, when $\eta < 1$ it favors stronger correlation, whereas when $\eta > 1$ it favors weaker correlation. For the random effects covariance matrix under the sender $\bm\Sigma_{\gamma}$, the same prior is specified.

Finally, multivariate normal priors are specified for the fixed effects and random effects, i.e.,
\begin{eqnarray*}
\bm\phi &\sim& N(\textbf{0},\bm\Psi_{\phi})\\
\bm\psi &\sim& N(\textbf{0},\bm\Psi_{\psi})\\
\bm\mu_{\beta} &\sim& N(\textbf{0},\bm\Psi_{\mu})\\
\bm\zeta_{\gamma} &\sim& N(\textbf{0},\bm\Psi_{\zeta}).
\end{eqnarray*}
For the prior covariance matrices, diagonal matrices are specified with very large variances. This builds upon existing knowledge of weakly informative prior specification for mixed effects generalized linear models, which have become the standard approach under this class of mixed effects models \citep{gelman2013bayesian, gelman2006data}.




\subsection{Reparameterization to improve Bayesian computation}\label{sec:reparm}

\par
Due to the hierarchical structure of the data, the random-effects parameters $\bm{\beta}_k$ and $\bm{\gamma}_{k}$ are highly correlated with the population parameters $\bm{\mu}$, $\bm{\zeta}$ and $\bm{\Sigma}$. This introduces severe computational inefficiencies of the sampling process. When the data are sparse, which is a characteristic of most social network data, the geometry of the posterior distribution makes it very difficult to sample from the highest posterior density areas. \cite{betancourt2015hamiltonian} called these issues \textit{pathologies of the hierarchical model}. Therefore, to ease the burden on the sampler, we take advantage of the multivariate normal structure of the random effects and apply a non-centered linear transformation to those parameters.

\begin{lemma}\label{theo1}

Let $\bm{\beta} \sim \mathcal{N}(\bm{\mu}, \bm{\Sigma})$, where $\bm{\beta} \in {\rm I\!R}^p$. Then, with $\bm{\mu} \in {\rm I\!R}^p $ and $\bm{A}$ being a $p \times p$ matrix, such that $\bm{A} \bm{A}' = \bm{\Sigma}$, one can write $\bm{\beta} = \bm{A} \bm{Z} + \bm{\mu}$, where $\bm{Z} \in {\rm I\!R}^p$ and $\bm{Z} \sim \mathcal{N}(\bm{0}, \bm{I})$, where $\bm{I}$ is a $p \times p$ identity matrix. 

\end{lemma}

\noindent
This transformation can be applied to both $\bm{\beta}$ and $\bm{\gamma}$. A natural candidate for matrix $\bm{A}$ is the Cholesky factorization of the covariance matrix $\bm{\Sigma}$. For details see appendix \ref{app:A}.

\par
This reparameterization is more efficient for two reasons. First, it reduces the dependency between the random effects parameters and the population parameters by sampling from independent standard normal distributions. This simplifies the geometry of the posterior and avoids inverting $\bm{\Sigma}$ at every evaluation of the multivariate normal density \citep{carpenter2017stan}. Therefore, we can safely and efficiently transform the random-effects parameters without causing any change in the prior specification of population parameters.

\par
The model has been implemented in Stan, a probabilistic programming language that employs Hamiltonial Monte Carlo (HMC) algorithms to sample from posterior distributions. The advantage of HMC methods is that they avoid the random walk behavior and the sensitivity to posterior correlations that plague many Bayesian applications \citep{hoffman2014no}, including hierarchical models. This particularities allow HMC to generally converge to high dimensional distributions much faster than Metropolis-Hastings or Gibbs sampler methods \citep{hoffman2014no, betancourt2015hamiltonian}. 
 
\section{Bayesian hypothesis testing under the actor-oriented multilevel relational event model}\label{sec:hyp_test}

\par
In multilevel analysis, researchers are usually interested in testing which theories receive the most support from the observed data, so that inferences about the population can be conducted. This kind of analysis is carried out through a process called hypothesis testing. From a Bayesian perspective those tests are usually performed by the computation of Bayes factors (\text{BF}) \citep{jeffreys1961theory}. The \text{BF} is given by the ratio of marginal likelihoods under the parameter space of competing hypotheses, which quantifies the probability of observing the data under one hypothesis relative to another hypotheses, and thereby providing a quantification of the relative evidence in the data between the hypotheses. Let
$\textbf{E}$ be the observed data, $\bm{\theta}$ a vector of parameters in the space $\bm{\Theta}$, and $\text{H}_{0} \in \bm{\Theta}_{0}$ be a hypothesis that will be tested against $\text{H}_{1}  \in \bm{\Theta}_{1}$, then $\text{BF}_{01}$ is expressed as
\begin{equation}
    \label{BF}
    \text{BF}_{01} = \frac{m(\textbf{E} | \text{H}_{0})}{m(\textbf{E} | \text{H}_{1})},
\end{equation}

\noindent
where $m(\textbf{E} | \text{H}_{i}) = \int_{\bm{\theta}_{i} \in \bm{\Theta}_{i}} p(\bm{E} | \bm{\theta}_{i}) p(\bm{\theta}_{i}) d\bm{\theta}_{i}$, for $i = 0, 1$, with $p(\bm{E} | \bm{\theta}_{i} )$ being the likelihood and $p(\bm{\theta}_{i})$ the prior. Also, $\bm{\Theta}_{0} \cap \bm{\Theta}_{1} = \varnothing$, with both $\bm{\Theta}_{0}$ and $\bm{\Theta}_{1}$ being subsets of $\bm{\Theta}$. \cite{kass1995bayes} provide a rule-of-thumb for Bayes Factors interpretation. In their setting, the evidence provided by the $\text{BF}_{01}$ in favor $\text{H}_{0}$ can be seen as ``insufficient" if $1 < \text{BF}_{01} < 3$, ``positive" if $\text{BF}_{01} > 3$, ``strong" if $\text{BF}_{01} > 20$, and ``very strong" if $\text{BF}_{01} > 150$. These are rough guidelines to aid the interpretation of Bayes factors and should not be used as strict cut-off values.

Furthermore, when prior probabilities of the hypotheses have been formulated before observing the data, the prior odds can be updated with the Bayes factor to obtain the posterior odds of the hypotheses, i.e.,
\[
\frac{P(H_0|\textbf{E})}{P(H_1|\textbf{E})}=
BF_{01} \times \frac{P(H_0)}{P(H_1)}.
\]
For example, when both hypotheses are equally likely a priori, i.e., $P(H_0)=P(H_1)=\frac{1}{2}$, the posterior probabilities of the hypotheses are given by $P(H_0|\textbf{E})=\frac{BF_{01}}{BF_{01}+1}$ and $P(H_1|\textbf{E})=\frac{1}{BF_{01}+1}$. These posterior probabilities quantify the probabilities that each of these hypotheses are true after observing the data (when assuming that one of these hypotheses is true). Because Bayes factors are consistent under mild conditions, the evidence for the true hypothesis goes to infinity and the posterior probability of the true hypothesis goes to 1 as the sample goes to infinity.

Below we introduce Bayes factors for three types of hypothesis testing problems for the multilevel relational event modeling literature. Test I is useful for testing the homogeneity of network effects across independent sequences. Test II is useful for testing the degree of heterogeneity of network effects across sequences using order constraints on random effects variances. Test III is useful to test equality and order constraints on fixed effects. Prior specification and numerical computation, which are important aspects when computing Bayes factors and posterior probabilities of the hypotheses, are discussed for each test separately.

\subsection{Test I: Testing for homogeneous social interaction behavior across sequences}\label{rnd_var}
When building multilevel relational event models, a central question is whether a driver of specific social interaction behavior (as quantified via the network effects) is constant over the sequences or whether it varies across the sequences. Testing this is important from a statistical point of view (i.e., to keep the model as parsimonious as possible, and thus to avoid an enormous overparameterization using different effects across all $K$ sequences) but also from a substantive point of view (i.e., to understand which drivers of social interaction behavior in relational event data are constant across sequences and which (highly) differ). Thereby, testing this contributes to a better understanding of the heterogeneity of social interaction behavior.

The hypothesis test for the $p$-th random network effect in the receiver model can be formulated as $H_0:\sigma_{\beta,p}=0$ versus $H_0:\sigma_{\beta,p}>0$, or equivalently as, $H_0:\beta_{p,1}=\ldots=\beta_{p,K}$ versus $H_1:\text{not $H_0$}$. The second formulation of the hypothesis is simpler as we avoid the need of testing whether the variance is equal to the boundary of 0, but instead we only need to test whether the random effects are equal across the $K$ clusters. Using the fact that the distribution of the random effects effectively serves as a prior for the random effects on the first level, we can simply compute a Savage-Dickey density ratio using the estimates of the random effects distribution from the data, i.e., 
\begin{equation*}
    \text{BF}_{01} = \frac{\pi(\xi_{1}, \dots, \xi_{K-1} = 0| \textbf{E})}{\pi(\xi_{1}, \dots, \xi_{K-1} = 0)},
\end{equation*}
where $\xi_{p,k}=\beta_{p,k}-\beta_{p,k-1}$, for $k=2,\ldots,K$, are the differences between the random effects.

To compute the posterior and prior densities at the null value, we use the fact that our actor-oriented multilevel relational event model can be written as two independent mixed effects Poisson regression models having specific forms (Appendix \ref{app:AO-Poisson}). Thus the two independent components of the model belong to the family of generalized linear mixed effects models, which is well understood in the statistical literature \citep[e.g.,][]{gelman2006data,burkner2017brms}. Moreover, the model results in unimodal posteriors when using weakly informative priors, as done in this paper. Additionally, we use the general property that posterior distributions converge to Gaussian distributions in well-behaved problems following large sample theory (Gelman et al., 2004, Ch. 4; Rafter, 1995; Kim \& Ibrahim, 2000). We illustrate the accuracy of this Gaussian approximation later in this paper. The Gaussian approximation of the posterior is then used to compute the numerator (then having an analytic expression). Moreover, the prior distribution, which is needed to compute the denominator, follows a multivariate Gaussian distribution by definition because it is the second level of our multilevel model, $\bm\beta_k\sim N(\bm\mu_{\bm\beta},\bm\Sigma_{\bm\beta})$, and thus the joint prior distribution of the contrasts ($\xi$) also follows a multivariate Gaussian distribution.

One interesting aspect of this Bayes factor is that the prior is fully determined
from the data, similar to empirical Bayesian estimation of hierarchical models.
This property is especially useful here as Bayes factors are known to be
sensitive to the choice of the prior. Note that empirical Bayesian approaches to obtain Bayes factors have been proposed for testing regression coefficients using $g$ priors \citep[e.g., see][and the references therein]{liang2008mixtures}, but (to our knowledge) not for testing variance components as we do here. The advantage of this approach is
that no external prior information is required and no other ad-hoc choices for prior distributions are needed. The outcome of the test can be used to quantify the relative evidence in the data of whether a hierarchical structure for the network effects is applicable or not given the observed data.

\subsection{Test II: Testing the degree of heterogeneity of network effects across sequences}\label{var_const}

\par
After establishing which network effects are heterogeneous across sequences (using the test from the previous subsection), it is useful to investigate the relative degree of heterogeneity of network effects across sequences, again with the goal to better understand the heterogeneity of social interaction behavior across independent relational event sequences. This comes down to testing whether a specific random effect varies more across sequences than another random effect, or, equivalently, whether one random effect variance is larger than another. When generalizing this further, we would test specific orderings of random effect variances \citep[see also][who consider testing order constraints on variances in a nonhierarchical setting]{boing2020bayes}. 

Two different tests are proposed for this purpose. First a confirmatory test is proposed of whether an anticipated ordering regarding the degree of heterogeneity of network effects across sequences is present:
\begin{eqnarray*}
\text{H}_{0}&:& \sigma_1 < \sigma_2 < \dots < \sigma_{P}\\
\text{H}_{1}&:& \text{not $H_0$}
\end{eqnarray*}

Second, an exploratory test is proposed to determine which ordering out of all $P!$ possible orderings receives most evidence from the data, i.e.,
\begin{eqnarray*}
\text{H}_{1}&:& \sigma_1 < \sigma_2 < \dots < \sigma_{P}\\
\text{H}_{2}&:& \sigma_2 < \sigma_1 < \dots < \sigma_{P}\\
\vdots& & \\
\text{H}_{P!}&:& \sigma_P < \sigma_{P-1} < \dots < \sigma_{1}
\end{eqnarray*}

Following the existing Bayesian literature on order constrained hypothesis testing \cite[e.g.][]{klugkist2005inequality}, we specify the prior under an order-constrained hypothesis as a truncated version of an unconstrained prior truncated in the order-constrained subspace. The Bayes factor of an order-constrained hypothesis against the unconstrained hypothesis$, H_u$, can then be expressed as the ratio of the posterior and prior probability that the constraints hold (Appendix \ref{app:B}):
\begin{equation}\label{ordered_var}
    \text{BF}_{1u} = \frac{P(\sigma_1 < \sigma_2 < \dots < \sigma_{{P}}| \textbf{E},H_u)}{P(\sigma_1 < \sigma_2 < \dots < \sigma_{{P}}|H_u)}.
\end{equation}
The Bayes factors between constrained hypotheses of interest can then be obtained using the transitive property of the Bayes factors, e.g., $B_{12}=B_{1u}/B_{2u}$.

The Bayes factor will not be sensitive to the prior as long as very vague identical priors are specified for the variances. In this case, the prior probability in the denominator will be equal to $(P!)^{-1}$, and the posterior probability will be fully determined by the information in the data (similar as in Bayesian estimation). Bartlett's paradox is thus not an issue in Bayesian order or one-sided hypothesis testing \citep{jeffreys1961theory,klugkist2007bayes,liang2008mixtures,mulder2014bayes}. 

To compute the posterior probability that the order constraints hold under the unconstrained model, we can simply fit the unconstrained model (using independent vague priors) and compute the proportion of draws satisfying the order constraints. The current test extends the use of Bayes factors for testing order hypotheses on variance components \cite[e.g.,][]{boing2017bayesian,mulder2019bayes,boing2020bayes} to multilevel relational event models.

\subsection{Test III: Testing common and average network effects over all sequences}\label{mean_rnd}

\par

In most applications, hypothesis tests are formulated on the common network parameters across clusters (i.e., the fixed effects) and on the average of the random-effects across clusters (i.e., the global random effects means). Since both parameters have the same role, hypothesis tests of these parameters are both discussed here using the same methodology.

First, a common test is of whether a network statistic has no effect, a negative effect, or a positive effect on the interaction behavior, which could be formulated as
\[
H_0:\psi_1=0 \text{ versus }
H_1:\psi_1<0 \text{ versus }
H_2:\psi_1>0
\]
for the first fixed effect of the receiver model, for instance. Second, hypotheses are often formulated on combinations of parameters using equality and/or order constraints on multiple parameters of interest based on existing theories or scientific expectations \citep{Hoijtink:2019}, e.g., 
\[
H_1:\psi_1=\psi_2<\psi_3 \text{ versus }
H_2: \text{ `not $H_1$',}
\]
where hypothesis $H_1$ assumes that the first effect ($\psi_1$) is equal to the second effect ($\psi_2$) and larger than the third effect ($\psi_3$), and hypothesis $H_2$ assumes the complement is true.
Third, when testing the effects of categorical (dummy) variables on the event rate, scientific expectations may be formulated which categorical variable has the largest impact on the event rate. A challenging testing problem is when the interest is in assessing which categorical variable has the largest impact when it is not known which category of each categorical variable results in the largest event rate. For example, it may be of interest whether the dichotomous gender variable (0 or 1) has a larger, equal, or smaller impact on the event rate than a dichotomous race variable (0 or 1), but it is not of particular interest which category results in the largest rate. This could be translated to constrained hypotheses on the absolute values of the effects of these two categorical variables according to  
\[
H_0:|\psi_1|=|\psi_2| \text{ versus }
H_1:|\psi_1|<|\psi_2| \text{ versus }
H_2:|\psi_1|>|\psi_2|.
\]
Finally, it is possible to test parameters between the sender model and the receiver model. This allows researchers to assess whether variable has a larger/equal/smaller to predict the next sender than to predict the next receiver.

The procedure for Test III builds on default Bayes factor methodology where the information in the data is split between a minimal subset, which is used for default prior specification (in combination with a noninformative prior) and a maximal subset which is used for hypothesis testing \citep[][among others]{o1995fractional,perez2002expected}. When using these methods, the resulting Bayes factors do not depend on the undefined normalizing constants of the improper priors \citep{o1995fractional}, which practically means that arbitrarily vague priors can be used, as we do in this paper. These Bayes factors can be computed in an automatic fashion and manual prior specification can be avoided \footnote{Alternatively, JZS priors \citep{bayarri2007extending, rouder2009bayesian,wetzels2012default} could also be considered for this test but to our knowledge this class of priors has not yet been proposed for generalized linear mixed effects models. Moreover, note that the prior scale of the key parameter that is tested still needs to be manually specified, unlike the proposed fractional Bayes factor using a minimally informative fractional prior.}. Here we use fractional Bayes factors where a (minimal) fraction of the data, denoted by $b$, is used to construct a (default) fractional prior, given by
\begin{equation}\pi(\bm{\phi},\bm{\psi},\bm{\mu},\bm{\zeta},\bm\Sigma_{\beta},\bm\Sigma_{\gamma} | \textbf{E}^b) \propto p(\textbf{E} | \bm{\phi},\bm{\psi},\bm{\mu},\bm{\zeta},\bm\Sigma_{\beta},\bm\Sigma_{\gamma})^b \pi(\bm{\phi},\bm{\psi},\bm{\mu},\bm{\zeta},\bm\Sigma_{\beta},\bm\Sigma_{\gamma}),
\end{equation}
and the remaining fraction is used for hypothesis testing. It has been shown that fractional Bayes factors follow many important properties which are not shared by all default Bayes factors, such as large sample consistency, satisfying the likelihood principle, invariance to transformations of the data \citep{o1997properties}, and information consistency \citep{mulder2014prior}. To properly incorporate the model complexity of order-constrained hypotheses, we apply a prior adjustment of the fractional prior \citep{mulder2014prior}, and Gaussian approximations are applied to the posterior and fractional prior to simply the computation (Kim \& Ibrahim, 2000; Gu, et al., 2017). The accuracy of the Gaussian approximations will be discussed in the next section. The approximated fractional Bayes factor of a constrained hypothesis (e.g., with any set of equality and/or constraints on certain parameters) against an unconstrained model is then computed as the integral over the unconstrained posterior over the constrained subspace divided by the integral over the unconstrained fractional prior over the constrained subspace as a Savage-Dickey density ratio \citep{mulder2014prior}. Existing functions in R using the `mvtnorm' package (Genz et al., 2021) can be used to compute the Bayes factors \citep{mulder2019bfpack}.

The choice of the fraction `$b$' is based on the recommendation by \cite{mulder2019bayes}, which implies selecting a minimal sample that is based on the ratio between the total number of parameters (excluding group specific effects) and the total number of observations. Since sender and receiver models are two different processes assumed to be conditionally independent given the past, we have to define two difference fractions, one for each model \citep*[see also][]{Hoijtink:2019}. In the sender model, we have $P$ random effects and $Q$ fixed effects, hence the fraction is $b_{\text{Snd}} = \frac{(P(P+1)/2) + (P+Q)}{\sum_{k = 1}^{K} N_{k}}$ and in the receiver model we have $V$ random effects and $U$ fixed effects, resulting in the fraction $b_{\text{Rec}} = \frac{(V(V+1)/2) + (V+U)}{\sum_{k = 1}^{K} N_{k}}$. 


\section{Estimating and testing classroom dynamics using the multilevel relational event model}\label{sec:applic}

\par
The data that are considered here were collected by \cite{mcfarland2001student} in a study to investigate student rebellion in the classroom. The data feature observations of interactions among high-school students in two different schools in the United States. For this illustration, we consider 15 independent classrooms from Magnet High School during the 1996-1997 school year. The student body of this high school can be considered academically homogeneous. The data were collected through classroom observations in which the 
conversations within the classroom were coded. In each of these fifteen classes, a teacher is present in addition to the students. The number of events (the number of times one person said something to another person) ranges from 86 to 628, and the number of persons (the students plus the teacher) ranges between 19 and 30 across the event sequences. The conversations happened in an orderly lecture-like fashion, so only one person was speaking at each time. The aim of this application is two-fold. At first, we present a sequential analysis that is used as proof of concept for the proposed model and statistical testing procedures. Secondly, we aim to provide insights into classroom dynamics by showing the evolution of network effects over time and extensively applying our tests to showcase multiple substantively interesting hypothesis tests. Below we first discuss the actor-oriented multilevel relational event model that we use to analyze the data.

\subsection{Model specification}

\par
We fit the hierarchical actor-oriented relational event model (described in Section \ref{sec:model}) to the data. Initially, all effects are considered random, so the need for a hierarchical structure can be evaluated using the Bayes factor presented in Section \ref{sec:hyp_test}. This first step will allow us to determine whether some effects can be treated as fixed and the model can hence be simplified. 

\par
We set the prior parameters to $\sigma^2_{\phi} = \sigma^2_{\mu_{\beta}} = \sigma^2_{\psi} = \sigma^2_{\zeta_{\gamma}} = \sigma_{\tau} = 10$, making the priors for the fixed and random effects relatively vague, and $\eta = 2$, slightly favoring smaller correlations. For our illustration of the proposed methods, we will include the following covariates into the model:

\paragraph{Teacher:} A dummy variable that indicates whether the actor is the teacher (one if the actor is the teacher and zero otherwise). 
  
\paragraph{Gender:} A dummy variable that indicates the gender of the actor (one if the actor is male and zero otherwise).
  
\paragraph{Race:} A dummy variable that indicates the race of the actor. \cite{mcfarland2001student} notes that 50 percent of the \textit{Magnet High} population is Caucasian. Hence, the variable is one if the actor is Caucasian and zero otherwise. 

\paragraph{Inertia:} Inertia captures the persistence of the communication, where past interaction is likely to be repeated \citep*{leenders2016once}. It is computed as the accumulated volume of past communication from a specific sender to a specific receiver. 
Because this statistic is dyadic, it will be included only in the receiver model. 

\paragraph{Participation Shifts:} These statistics are used to reflect expectations of adherence of communication norms in small groups \citep{butts2008}. \cite{gibson2005taking} describes the framework for several types of participation shifts. In this application two distinct types of participation shifts will be included. The first belongs to the group of ``\textit{turn-receiving}" and is represented by the event pattern \textbf{ABBA}: an interaction from person A to person B is immediately followed by an interaction from B to A. The second is \textbf{ABAB}, which can be considered as a special case of "\textit{turn-continuing}": an interaction from person A to B is immediately followed by another interaction event from A to B. 

\par
The ABBA and ABAB participation shifts are concerned with dyads, therefore they will be included in the receiver model only. However, we do include adapted versions in the sender model. Here, the statistics become \textbf{ABB} (after A has spoken to B, the next event is B starting the conversation) and \textbf{ABA} (after A has spoken to B, A speaks again).

\paragraph{Activity:} This set of covariates captures the effect of actor activity as a sender or as a receiver \citep{vu2017relational}. The first is the  \textbf{\textit{outgoingness}} of a person, defined as the number of events sent by one actor up until a specific point in time. This captures the tendency of the person to start conversations (or just to talk). The second is the \textbf{\textit{popularity}} of a person, given by the number of events received by one actor up until a specific point in time. This captures the popularity of an individual as a receiver of the conversation.

\par
Relational event models are vulnerable to process explosion, which happens when $\lambda(t) \rightarrow \infty$ (often due to a feedback loop that may be caused by using statistics that are computed as cumulative sums \citep*{aalen2008survival}. This is particularly a problem for inertia and activity statistics. One way to alleviate this problem is via $z$-score standardization at every time point, which is defined as $z(t) = (x(t) - \bar{x}(t))/S_{x}(t)$, where $x(t)$ is the value of the statistic, $\bar{x}(t)$ is the sample mean, and $S_{x}(t)$ is the sample standard deviation at time $t$. We use standardized statistics in this application.

\subsection{Exploring social interaction dynamics as class time progresses
}


\par
By analyzing the relational event sequences as the time during class progresses, we can explore how statistical certainty increases over time using the proposed multilevel relational event model, and how interaction dynamics between children and the teacher evolves as class time progresses. To study this, we analyze the sequences with increasing batches of 20\% of the total number of events in each sequence (i.e. 20\%, 40\%, 60\%, 80\%, and 100\%).

\subsubsection{Testing for homogeneous social interaction behavior across school classes}

\par
\begin{figure}[t]

\subfloat{\includegraphics[width = 2in]{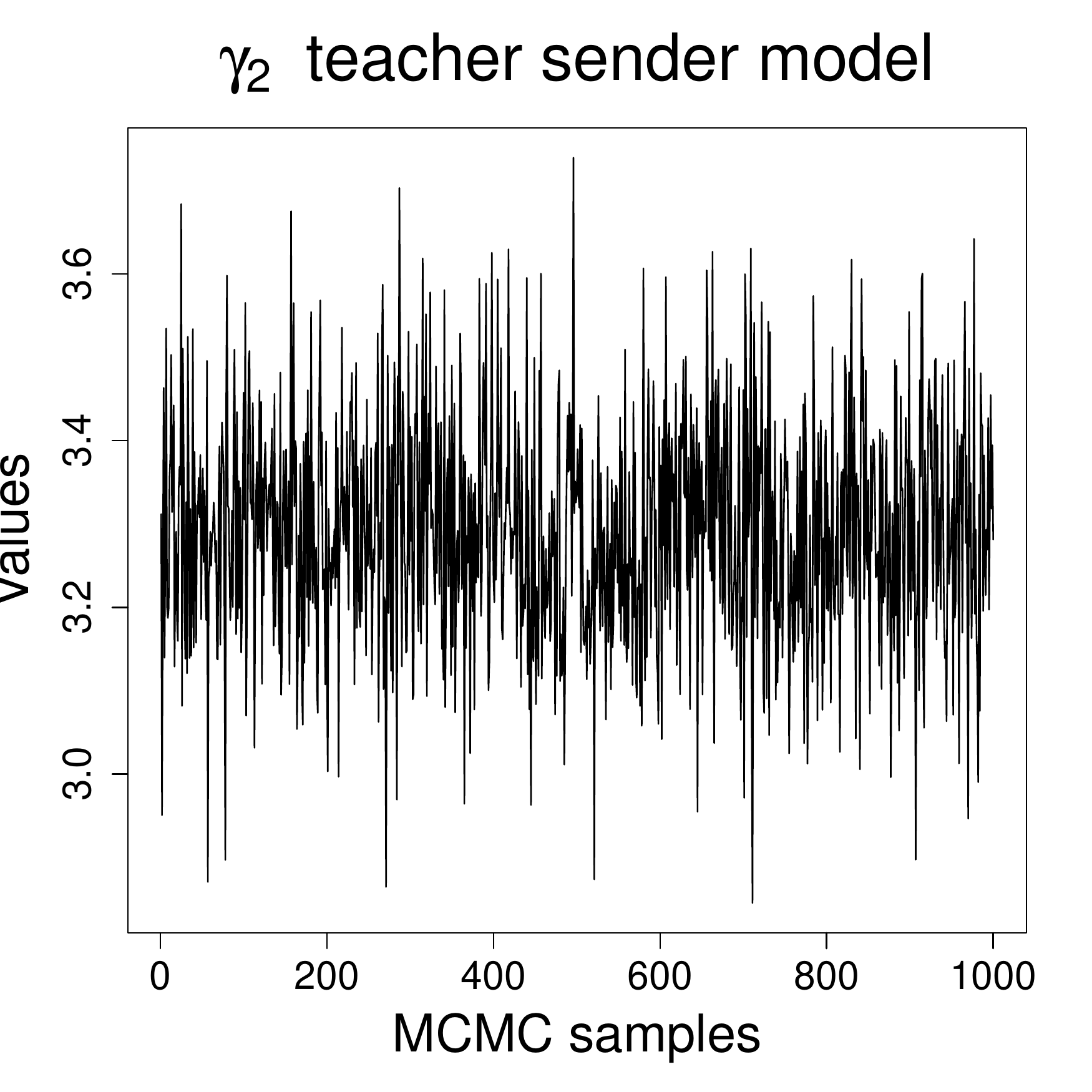}} 
\subfloat{\includegraphics[width = 2in]{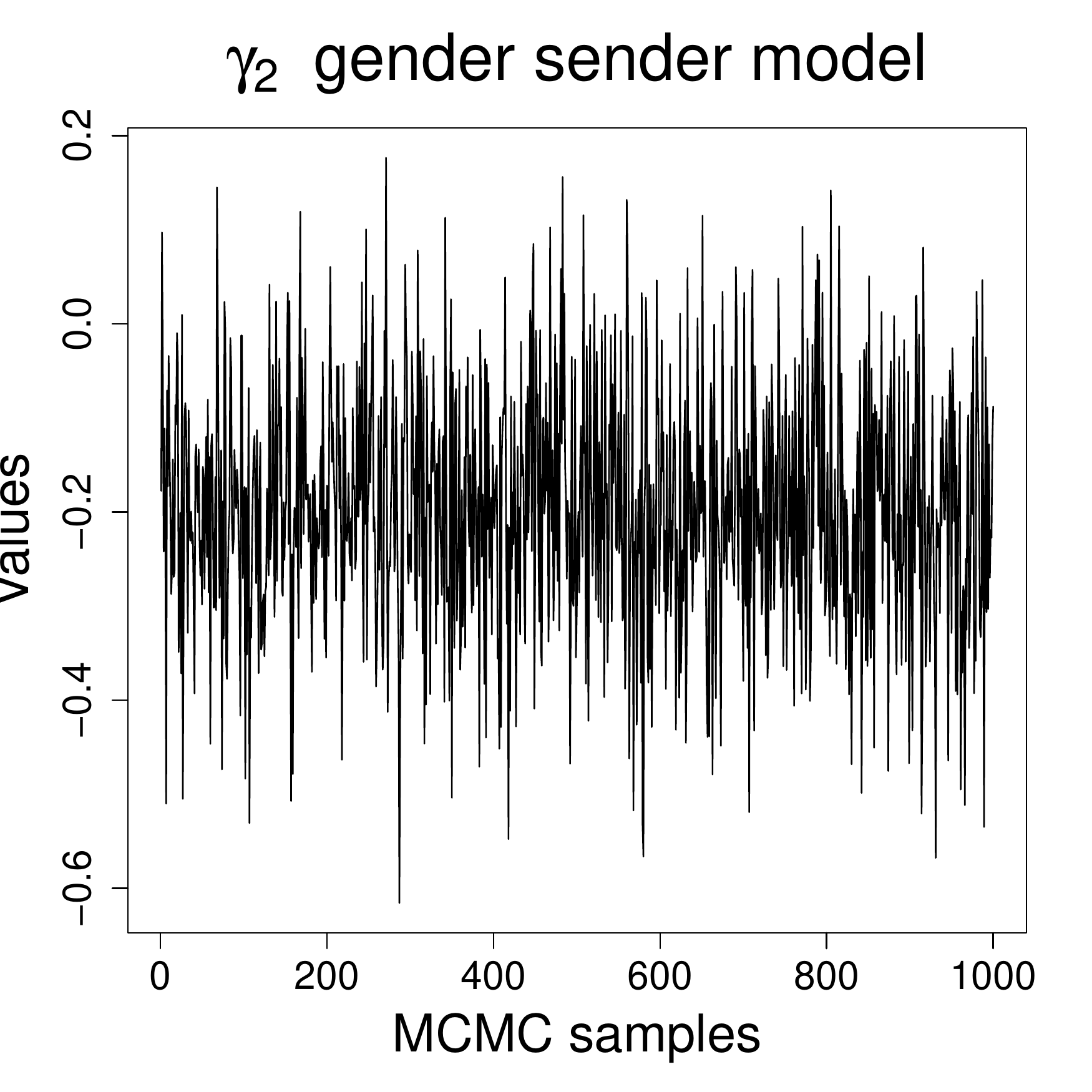}}
\subfloat{\includegraphics[width = 2in]{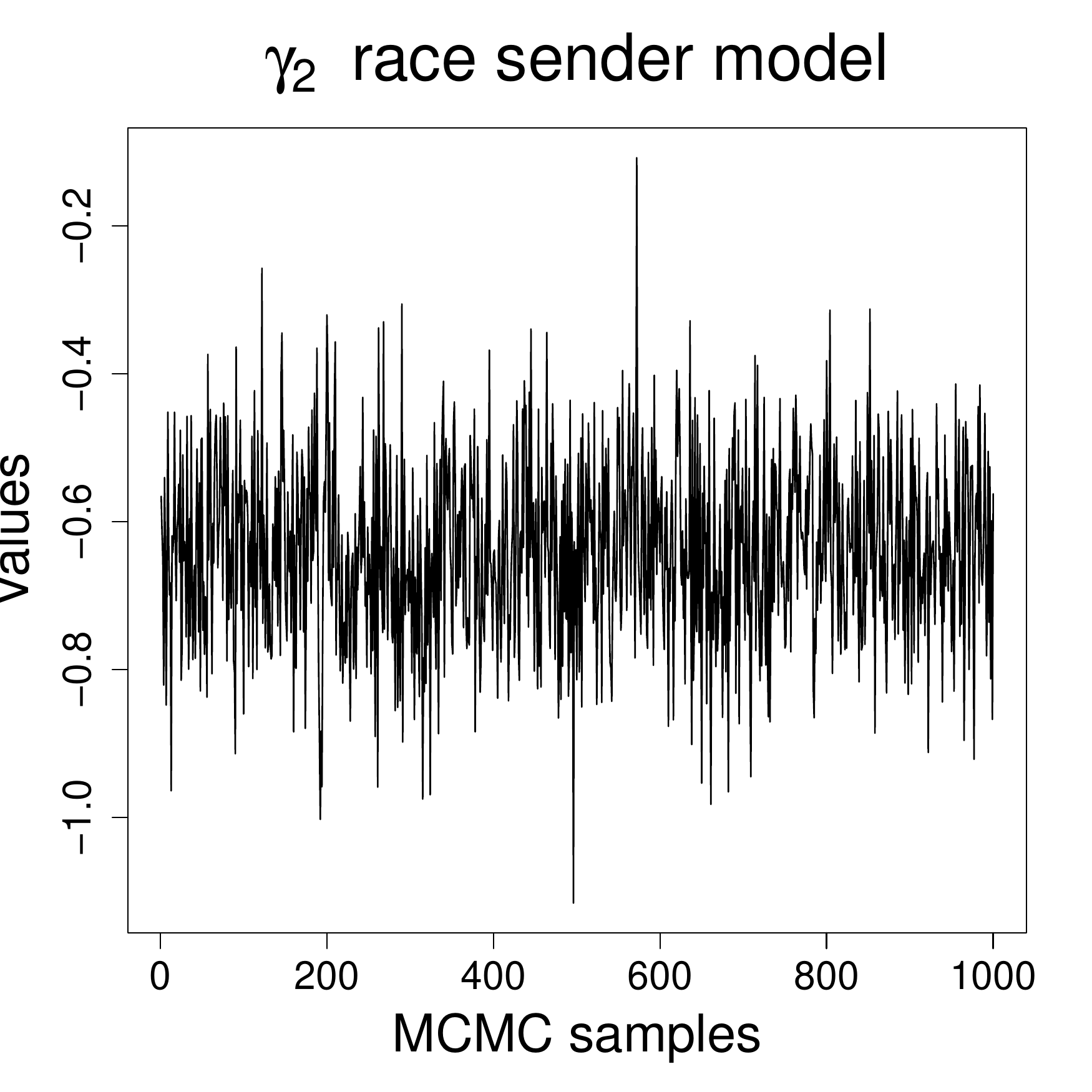}}

\subfloat{\includegraphics[width = 2in]{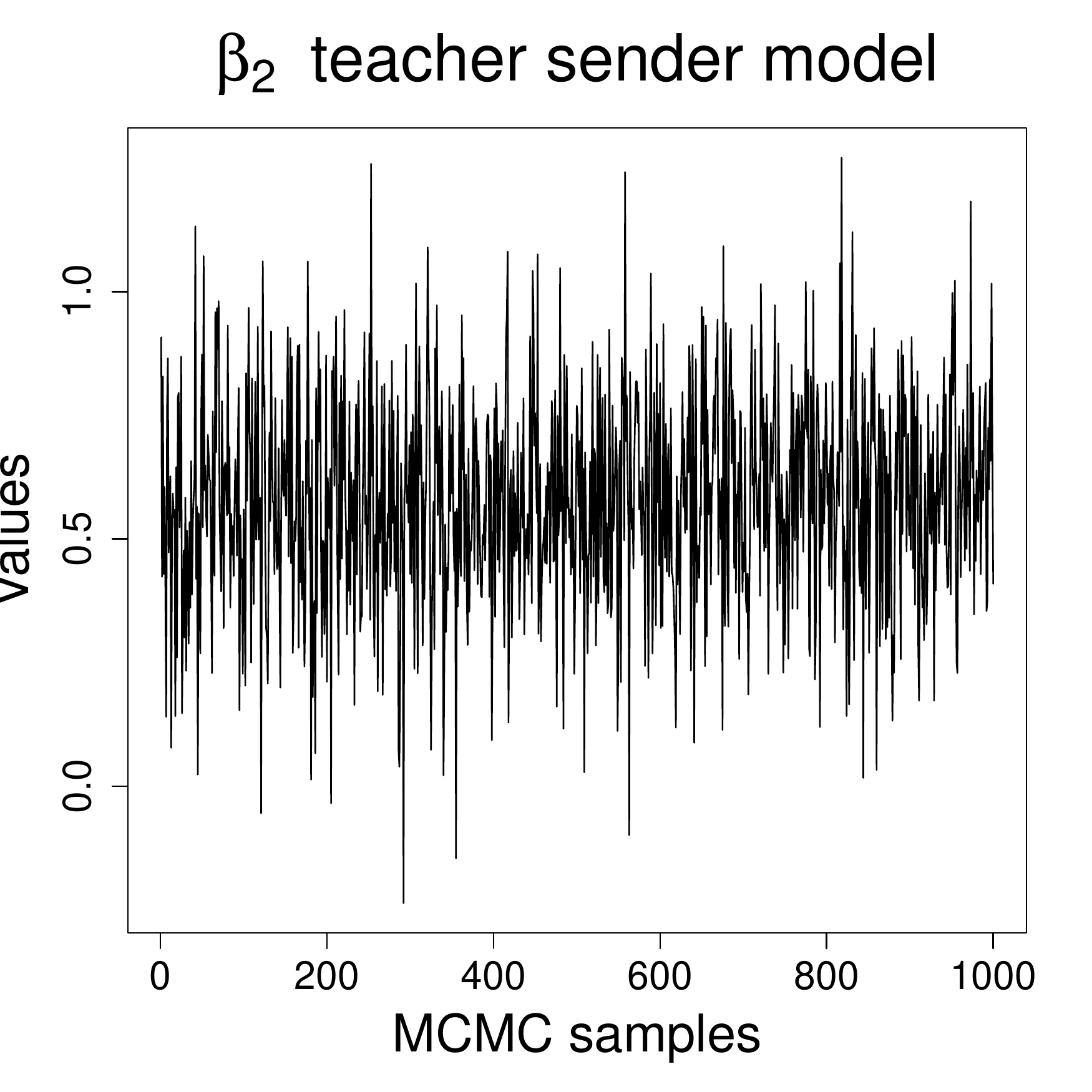}}
\subfloat{\includegraphics[width = 2in]{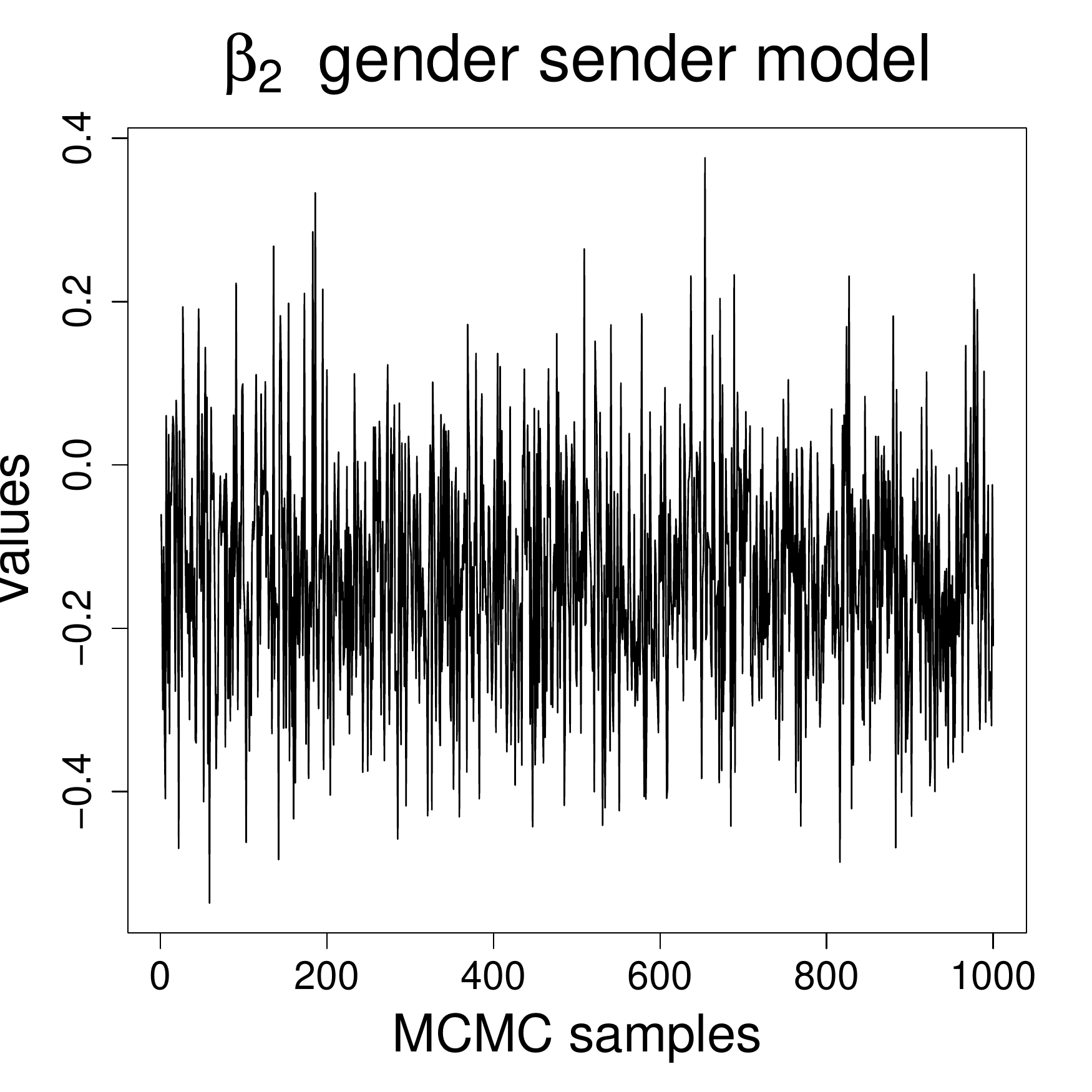}} 
\subfloat{\includegraphics[width = 2in]{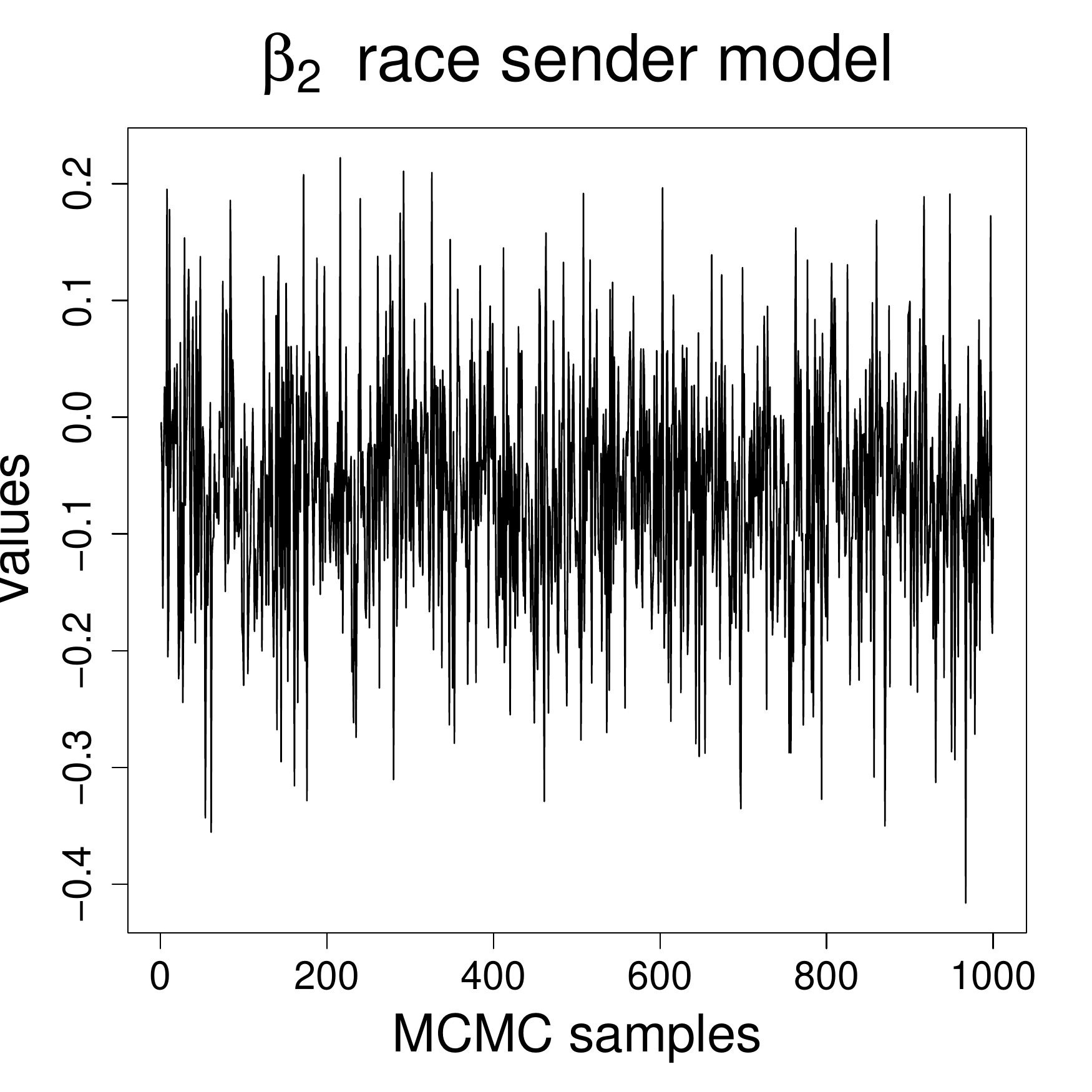}} 

\caption{Trace plots of some of the random-effect parameters. The number $2$ in the subscript shows that the parameters belong to cluster number two. The top (bottom) line shows parameters in the sender (receiver) model.}
\label{fig:trace_plots_rnd}

\end{figure}

The first step is to fit the model with all effects considered random. We ran $2000$ MCMC iterations and discarded the first $1000$ samples as burn-in. We used $\hat{R}$ as a convergence diagnostic measure. For all models, the metric values were fairly close to 1 all, falling all below the 1.05 value recommended by \cite{carpenter2017stan}. As an example, Figure \ref{fig:trace_plots_rnd} shows several trace plots for the model run with 100\% of the events. The chains show convergence. The number of posterior draws is not high because the Hamiltonian dynamics in Stan's algorithm allow for faster convergence than standard MCMC methods \citep{hoffman2014no}. To check this we investigated the bulk effective sample size and the tail effective sample size, which fell between 800 and 1100 for all parameters (\cite{carpenter2017stan} recommends these values to be both above 100 for the sample to be considered reliable). This confirms the computational efficiency of the sampler. For other sample size recommendations, see \cite{hecht2021sample}. Since a more parsimonious model is preferred, we consider an effect to vary across classrooms via a random effect when the posterior probability for the hypothesis that an effect is fixed, $\text{H}_{0}$, is less than 0.25, corresponding to $\text{BF}_{01} < 1/3$ (which can be interpreted as positive evidence that the effect varies across clusters; Kass \& Raftery, 1995). 

\par
\begin{figure}[t]
    \centering
    \subfloat{\includegraphics[width = 3.3in]{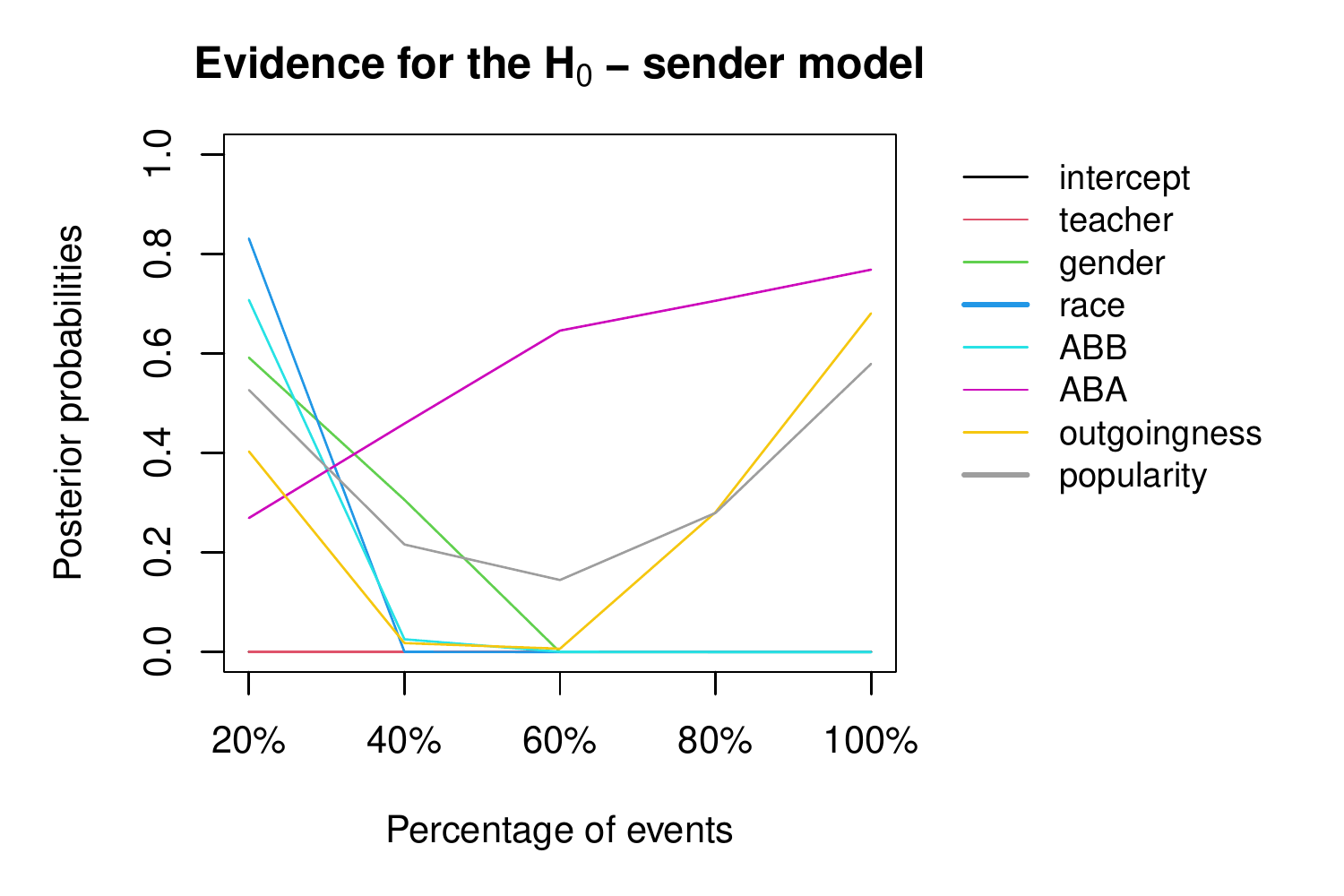}}
    \subfloat{\includegraphics[width = 3.3in]{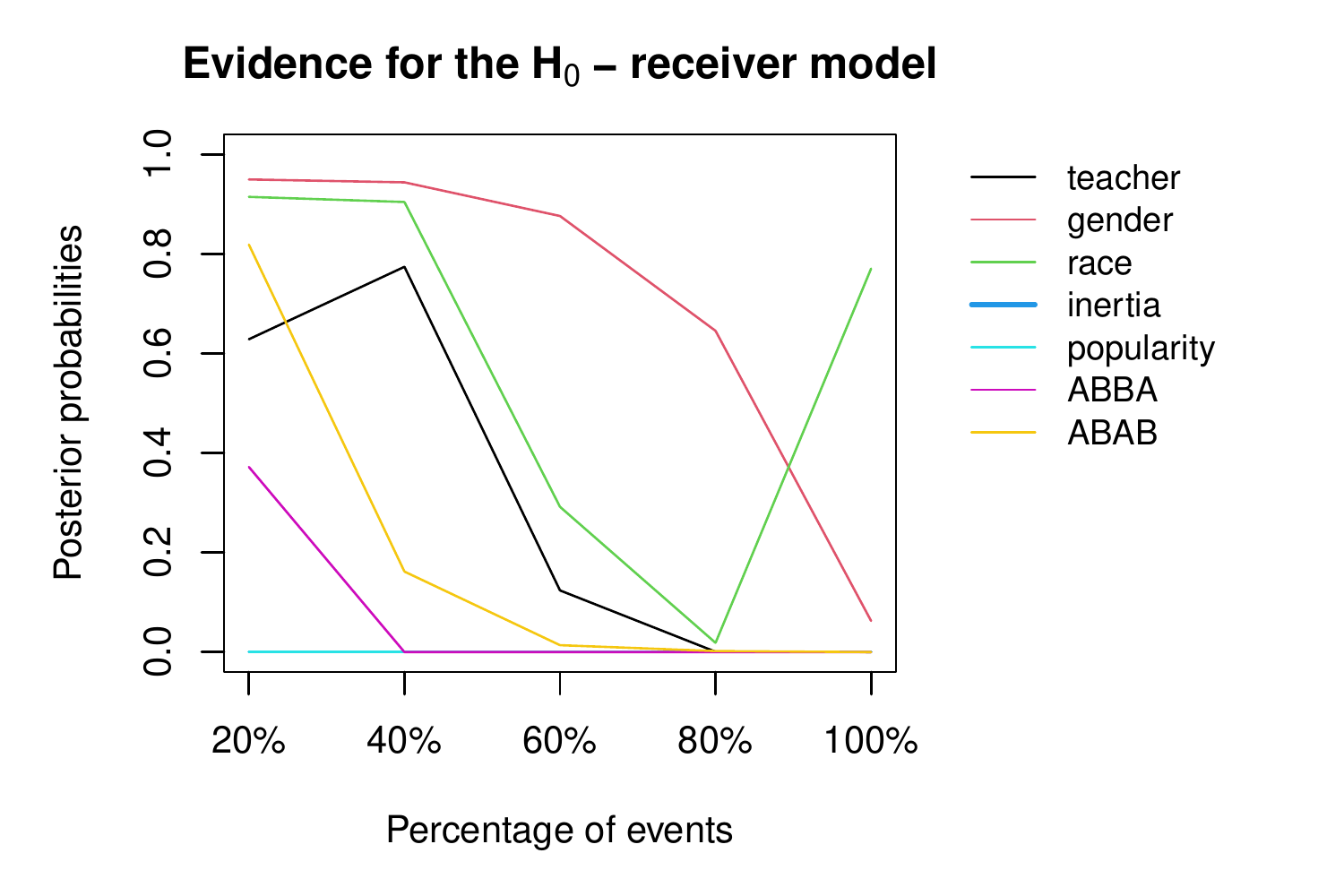}}
    \caption{Posterior probabilities of the fixed effect hypothesis $\text{H}_{0}$ in the homogeneity test for different effects.}
    \label{fig:homogeneity_test}
\end{figure}

Figure \ref{fig:homogeneity_test} shows the evolution of the posterior probabilities of $\text{H}_{0}$ as we increase the number of events. In the beginning, when the number of events is small (varying from 17 to 125 events across school classes), the uncertainty in the posterior is relatively large leading to considerable overlap in the posterior distributions of the random effects. Therefore, the posterior probabilities for a fixed effect tend to start out large and eventually decrease to zero when the full sequences are considered for most effects, providing strong evidence for their heterogeneous nature. This is the case for most effects that are clearly heterogeneous if the sample sizes are large enough.  On the other hand, for some covariates the posterior probability gets larger as the sample sizes near 100\% of the events. This is the case for \textbf{ABA}, \textit{\textbf{outgoingness}} and \textit{\textbf{popularity}} in the sender model and for \textbf{\textit{race}} in the receiver model, resulting in posterior probabilities for a fixed effect of 0.768, 0.681 and 0.579, 0.777, respectively. Thus, the parameters corresponding to those effects will be fixed in our model. Therefore, we continue with a mixed-effect model, where the sender model has \textbf{ABA}, \textbf{\textit{outgoingness}}, and \textbf{\textit{popularity}} as fixed effects, and  \textbf{\textit{intercept}}, \textbf{\textit{teacher}}, \textbf{\textit{gender}}, \textbf{\textit{race}}, and \textbf{ABB} as random effects, and a receiver model where only \textbf{\textit{race}} is a fixed effect and all other effects are considered random. 

\par
\begin{figure}[t]
\centering
    \subfloat{\includegraphics[width = 2.5in]{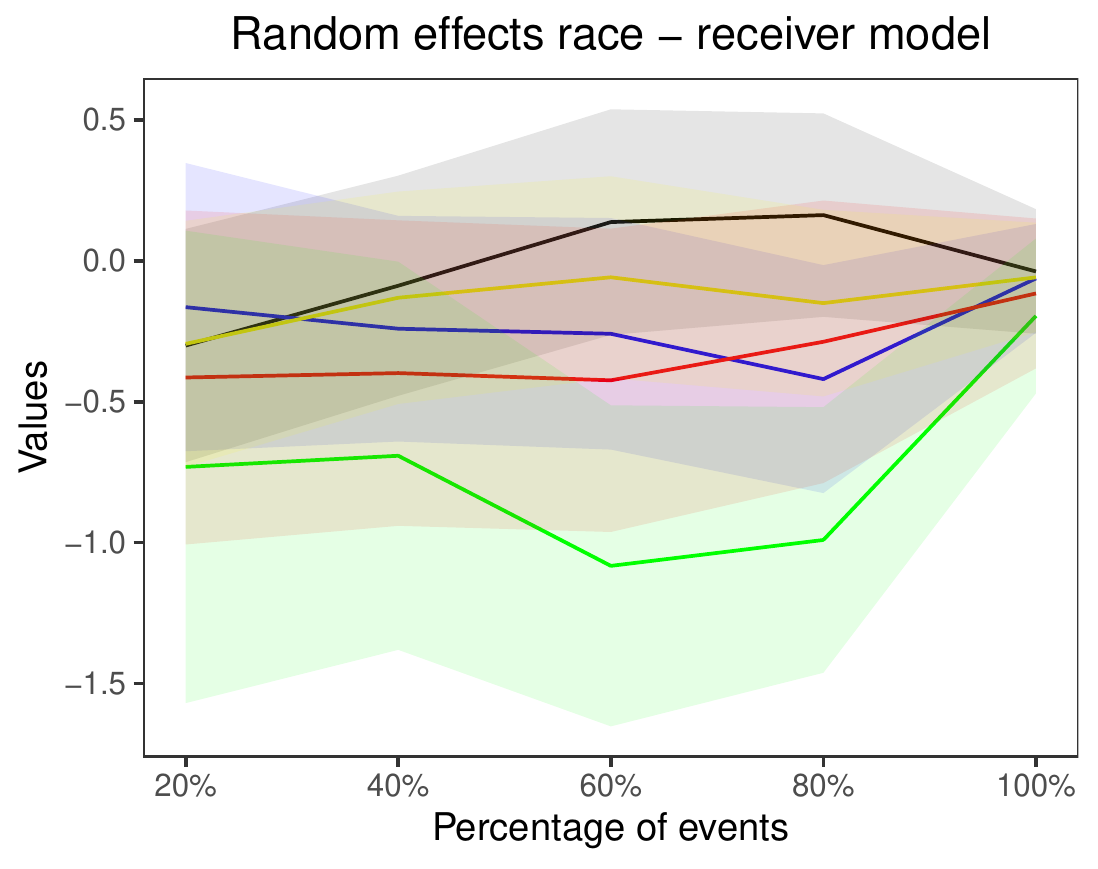}}
    \subfloat{\includegraphics[width = 2.5in]{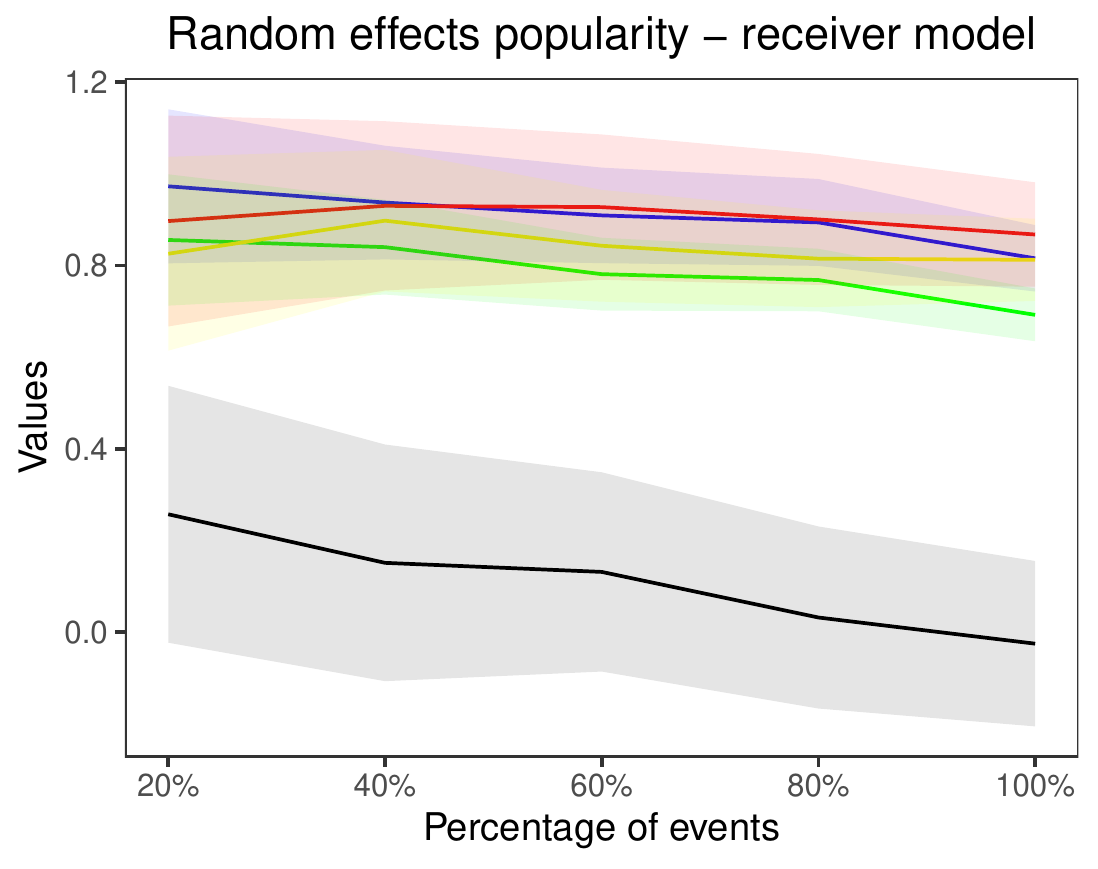}}\\
    \subfloat{\includegraphics[width = 2.5in]{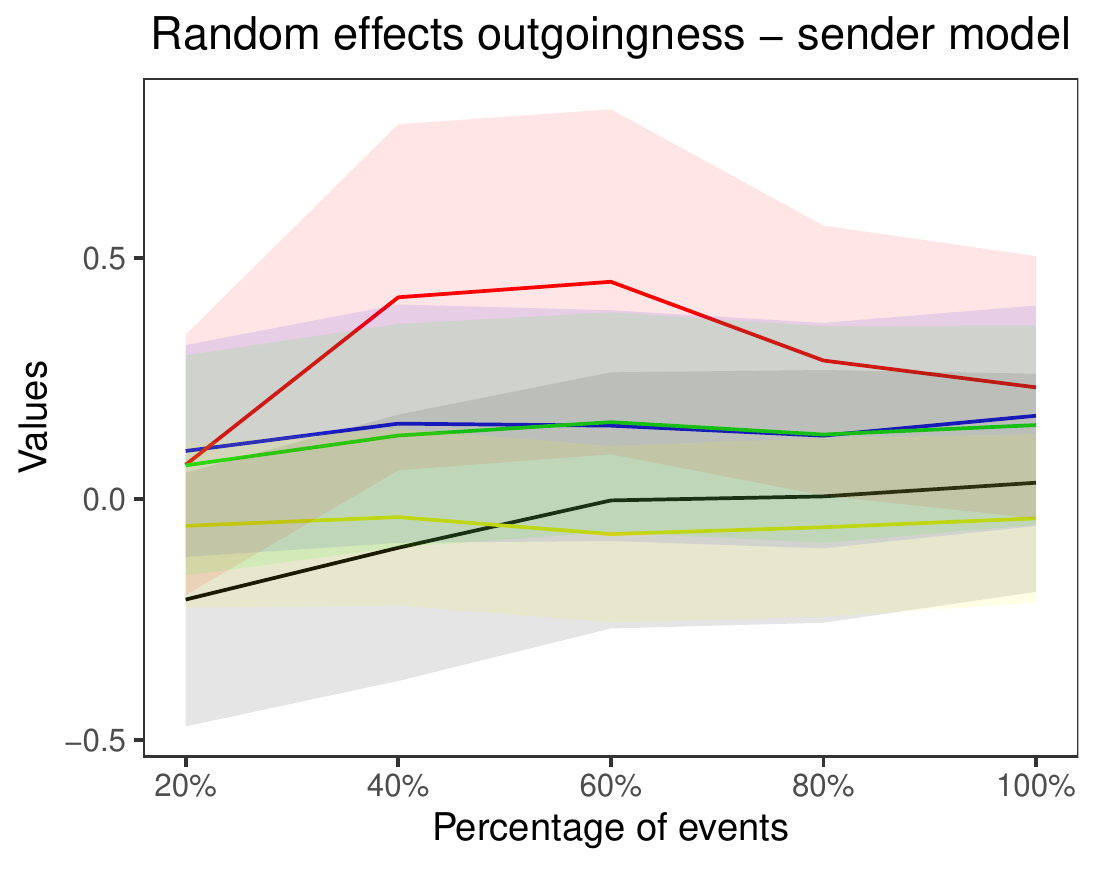}} 
    \subfloat{\includegraphics[width = 2.5in]{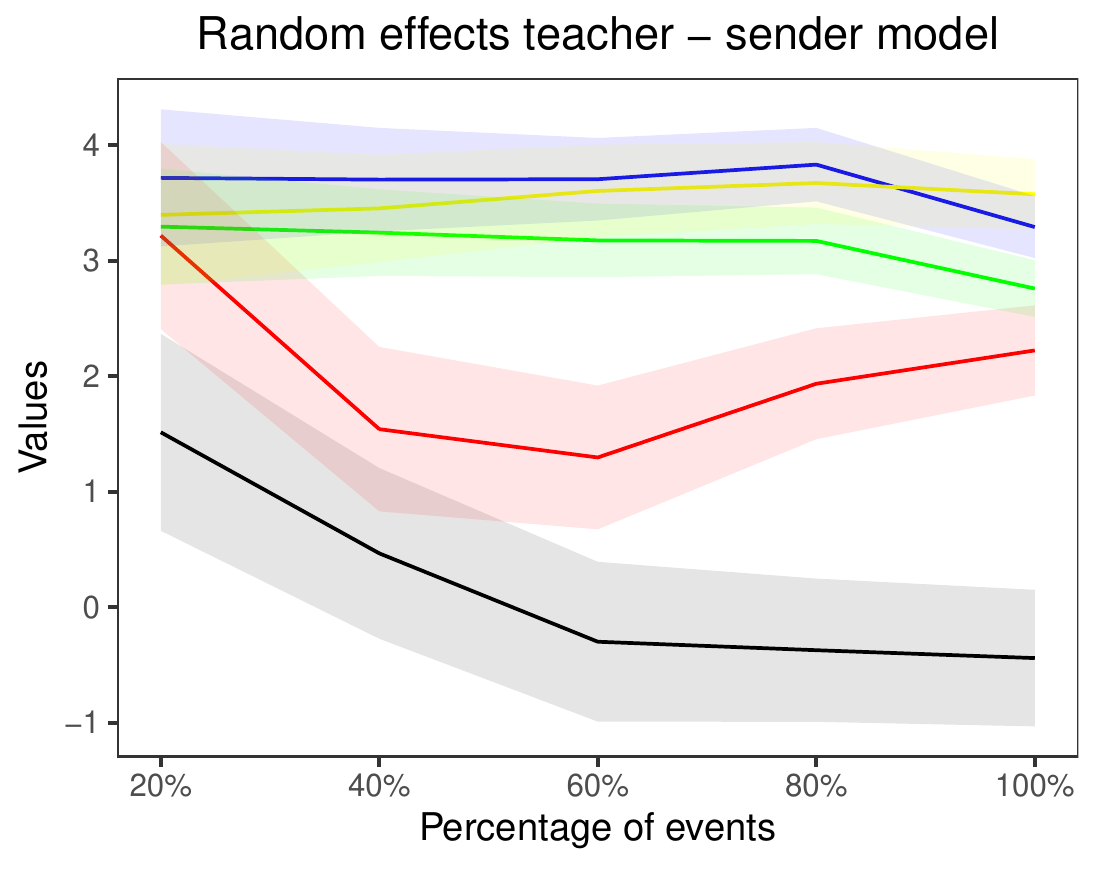}} 
    \caption{Random effect estimates for receiver (top) and sender (bottom) models for 5 groups.}
    \label{fig:random_effects}
\end{figure}

To understand the nonmonotonic development of the posterior probabilities where the lines first go down and eventually increase, it is useful to check the corresponding estimates of the random-effects. As an example, the top left panel of Figure \ref{fig:random_effects} shows random-effect estimates of the \textbf{\textit{race}} covariate in the receiver model, where each line represents a five different classrooms (to keep the plot clear; the other classrooms showed similar patterns). As the lines show we see that the random effects become more heterogeneous until 80\% of the events are included in the analysis but when all 100\% of the events are included the estimates become more homogeneous which explains the evidence for a fixed effect for the race effect in the receiver model in Figure 2 when all 100\% of the events are considered. 

\par
In addition, in the bottom right panel of Figure \ref{fig:random_effects}, which presents the random-effect estimates for \textbf{\textit{outgoingness}} in the sender model, a similar pattern is observed. The random effects becomes heterogeneous between 40\% and 60\% of the events, but they rapidly converge displaying a pattern similar to \textbf{\textit{race}} in the sender model. This behavior is reflected in the posterior probabilities, with the line going down for 40\% and 60\% of the events and then getting back up again. On the other hand, the estimates on the left hand side panels of Figure \ref{fig:random_effects}, which display \textbf{\textit{popularity}} in the receiver model and \textbf{\textit{teacher}} in the sender model, are consistently heterogeneous as class time progresses. This is corroborated by the posterior probabilities with both effects presenting small evidence for the null.  

\subsubsection{Evaluating the shrinkage behavior of random effects}

\par
Given that there is asymmetry in the distribution of the number of events across classrooms (varying from 86 to 628 events over classrooms), there will be a variation regarding the statistical uncertainty of the estimated random effects across classrooms. Due to the multilevel structure of the model, estimated effects with a relatively large uncertainty which also deviate much from the other estimates will then be (slightly) pulled towards the global mean of the effect over all class rooms. This statistical behavior is also known as shrinkage, or borrowing information across groups, which is an intrinsic property of multilevel models. The current sequential analysis where the sample sizes grows allow us 
to investigate this behavior, since with fewer events we should have less information and, therefore, less accurate classroom specific estimates. 

\par
\begin{figure}[t]
    \centering
    \includegraphics[width=5.5in]{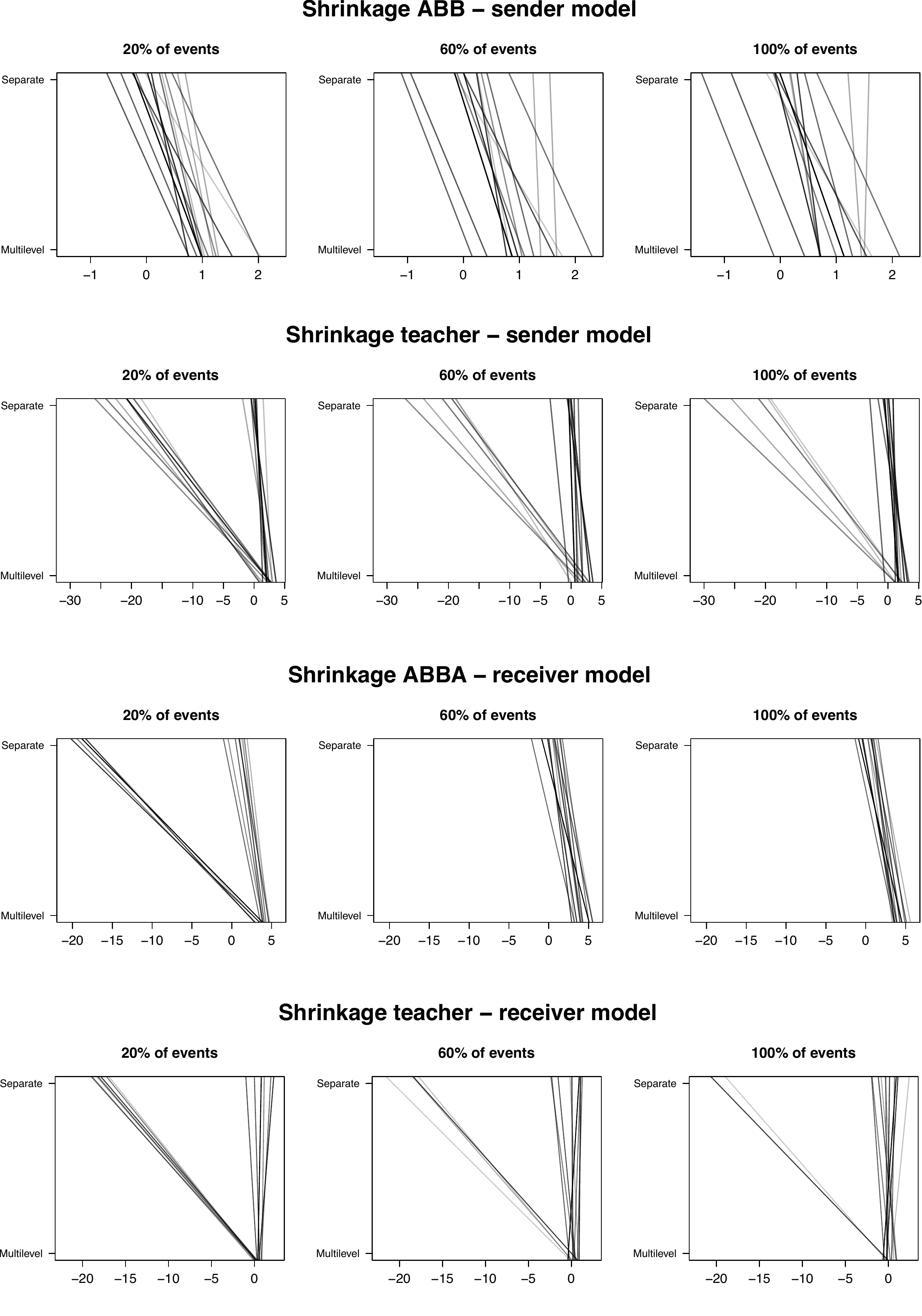}
    \caption{Plots of classroom-specific effects when fitting a model separately per classroom and when jointly fitting the multilevel model over all classrooms. Darker (lighter) lines indicate classrooms with relatively many (few) events.}
    \label{fig:shrinkage}
\end{figure}

We use the R package remstimate \citep{Arena2022} to obtain the estimates of the classroom specific effects which ignores the multilevel structure of the data.
Figures \ref{fig:shrinkage} shows the shrinkage for several effects in the sender and receiver models. The lines were plotted using a grey scale where lighter shades mean smaller sample sizes. In each panel, we have the estimates based on the estimation of the separate classrooms at the top, and the multilevel estimates at the bottom.

First, we observe see that there is considerably more variation of the estimated effects in the separate analyses (top of each panel), which ignore the multilevel structure, in comparison to the multilevel estimates (bottom of each panel). This confirms that the proposed multilevel model shows the anticipated shrinkage behavior. Second, we see that when using subsets of only 20\% complete event sequences in the classrooms we obtain more extreme estimates than when using the proposed multilevel model. Finally we see that estimates for classrooms with shorter event sequences (represented with light grey lines) are on average more extreme than estimates for classrooms for longer event sequences. This illustrates that the proposed multilevel relational event models is particularly useful when the event sequences are relatively short to avoid extreme estimates.


\subsubsection{Evolution of classroom behavior}

\par
\begin{figure}[t]
    \centering
    \includegraphics[width=6.0in]{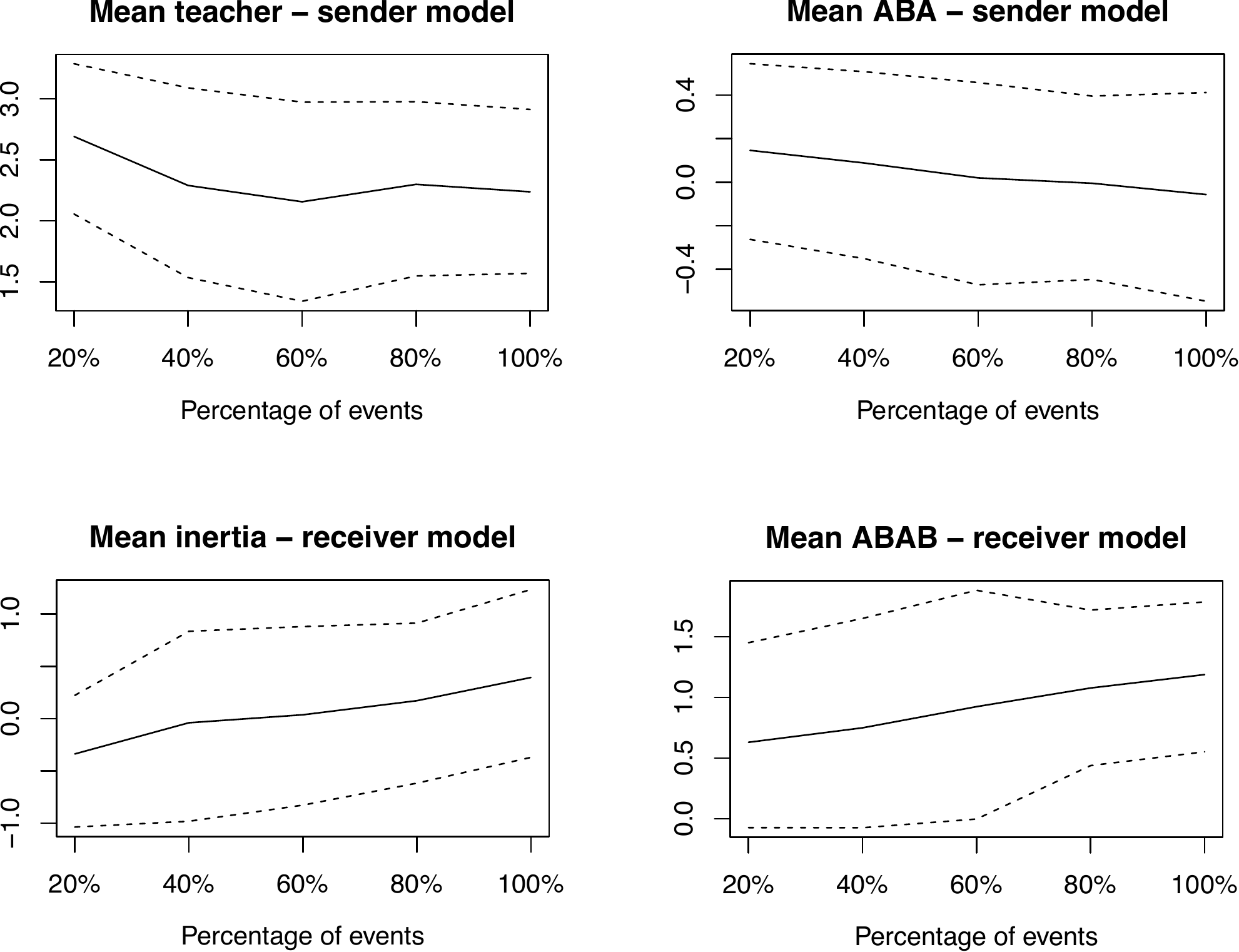}
    \caption{The development of different drivers of interaction behavior as class time progresses from 20\% to 100\% of the observed events in each classroom.}
    \label{fig:mean_trends}
\end{figure}

Another interesting aspect to look at in the sequential analysis is the potential change of the estimated of the effects as class time progresses. This will give us an insight into the social interaction dynamics in classrooms. Figure \ref{fig:mean_trends} shows the evolution of some mean effects in the sender and receiver models. Here we restrict ourselves to some trends of average network behavior across all classrooms. Note that the intervals are relatively wide because the limited lengths of the event sequences in a class period.

First, we see that the teacher effect in the sender model is positive and large in the sender model. This was expected given the dominant role of the teacher during lectures. The plot in Figure \ref{fig:mean_trends} (upper left panel) also shows a slight decrease as class time progresses. This suggests that the dominant role of the teacher slightly decreases as the lectures approach the end. This indicates that teachers may loose some of the control towards the end or that more students become engaged or activated towards the end. This is also confirmed by the ABA effect in the sender model (Figure \ref{fig:mean_trends}, upper right panel)) which is initially positive and decrease to just below zero towards the end of the class period. Thus, it becomes less likely that the sender of the previous messages (typically the teacher because these are lectures) becomes the sender of the next event.

Another interesting trend can be observed for the global inertia effect across classrooms in the receiver model (Figure \ref{fig:mean_trends}, lower left panel) In the beginning of the class period, inertia starts out negative which indicates that it is less likely that actors who were the receiver of relational events in the past (typically a question or remark by the teacher towards a student because lectures are considered) become a receiver of the next. This indicates that teachers aim to address different students in the beginning of the lecture with the goal to get more students engaged or activated during lectures. The effect however gradually increases as the class time progressed. As the class period nears the end, the inertia effect becomes positive which implies that it has become more likely that a student is addressed by the teacher who was also addressed by the teacher in the past. This suggests that teachers tend to focus more on the same students (possibly the more motivated students) rather than trying to involve other less active students in the discussion. This is also confirmed by the positive and increasing ABAB effect in the receiver model during class time (Figure \ref{fig:mean_trends}, lower right panel).



\subsection{Mixed-effects analysis on the full sequences}\label{sec:results}

\par
\begin{figure}[t]

\subfloat{\includegraphics[width = 2in]{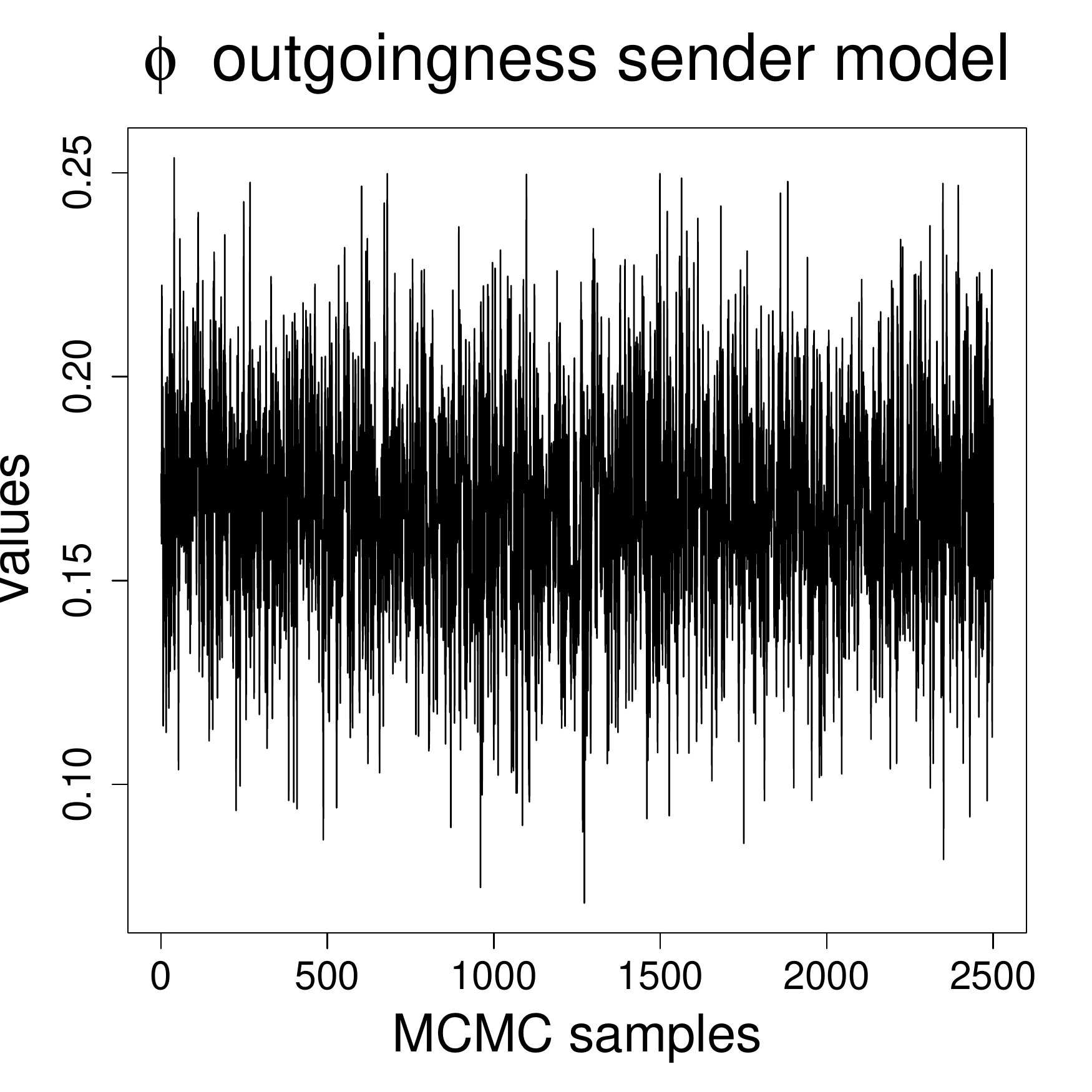}} 
\subfloat{\includegraphics[width = 2in]{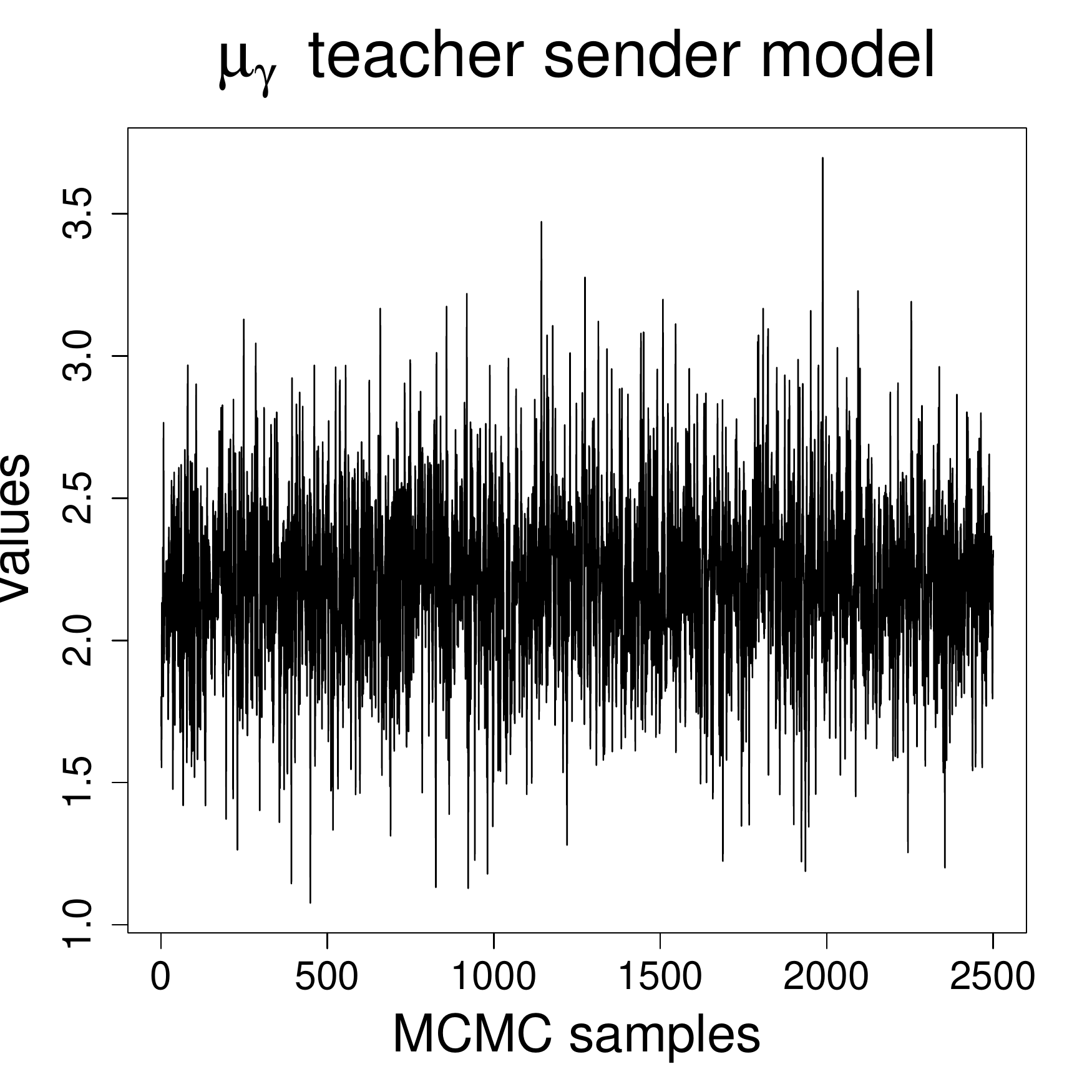}}
\subfloat{\includegraphics[width = 2in]{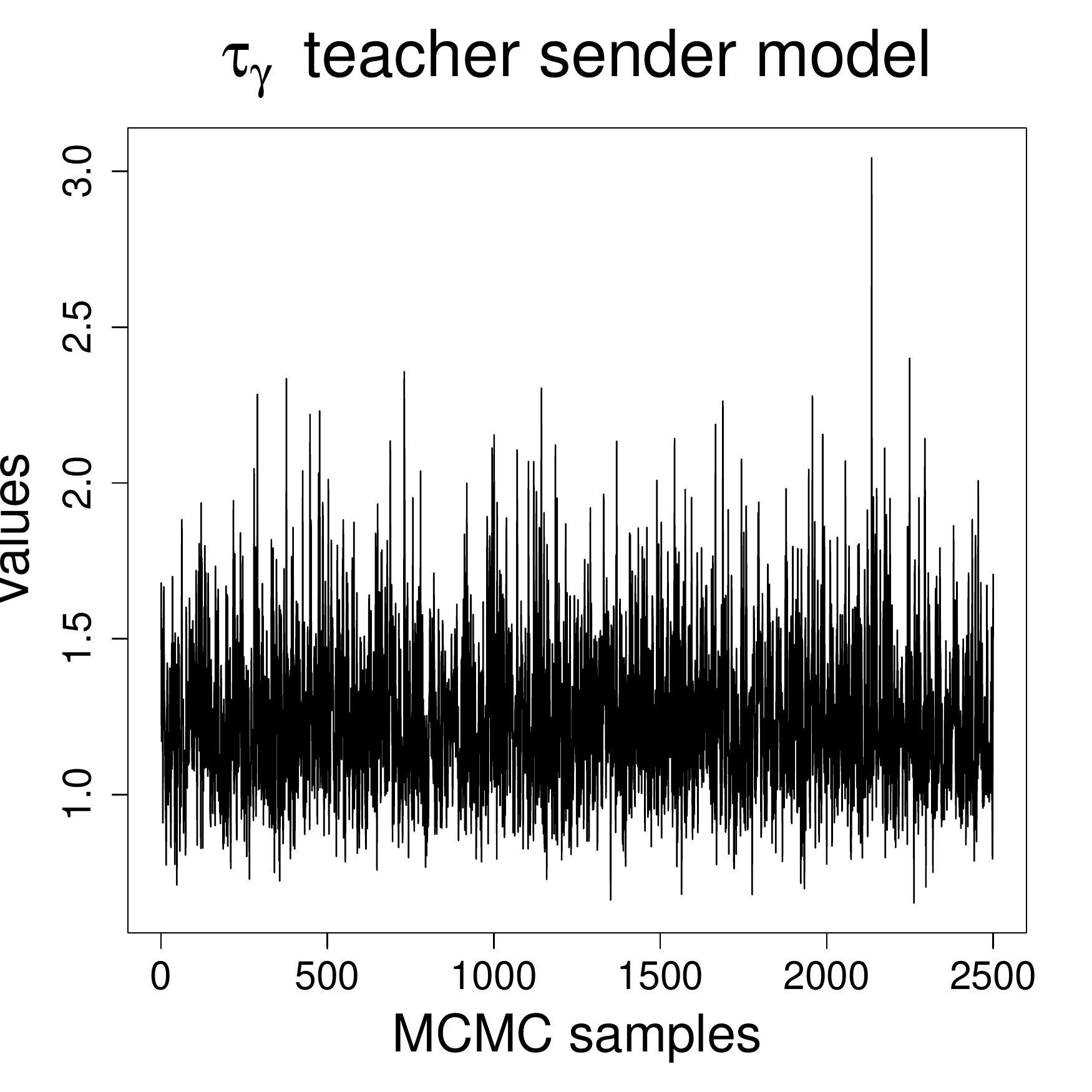}}\\

\subfloat{\includegraphics[width = 2in]{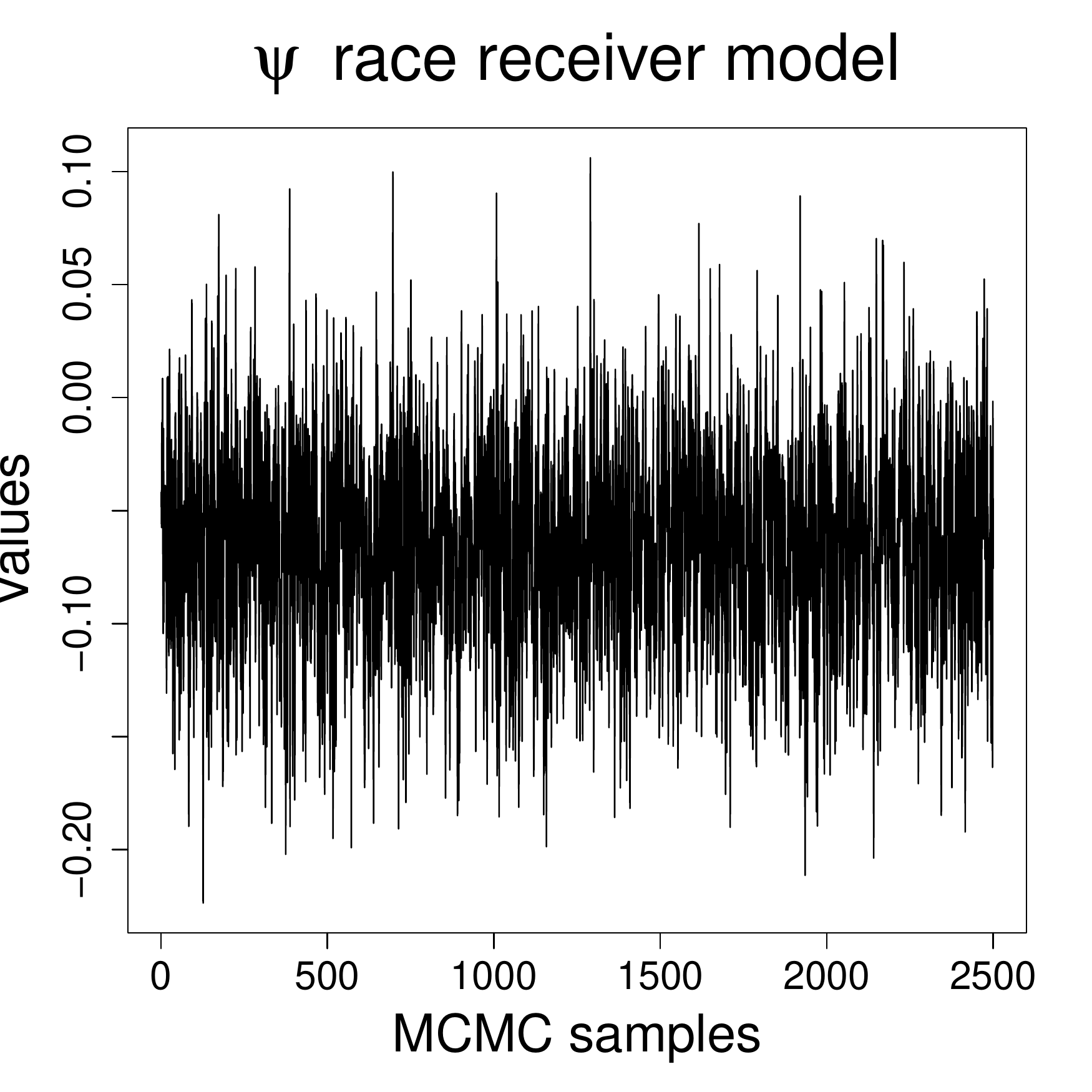}}
\subfloat{\includegraphics[width = 2in]{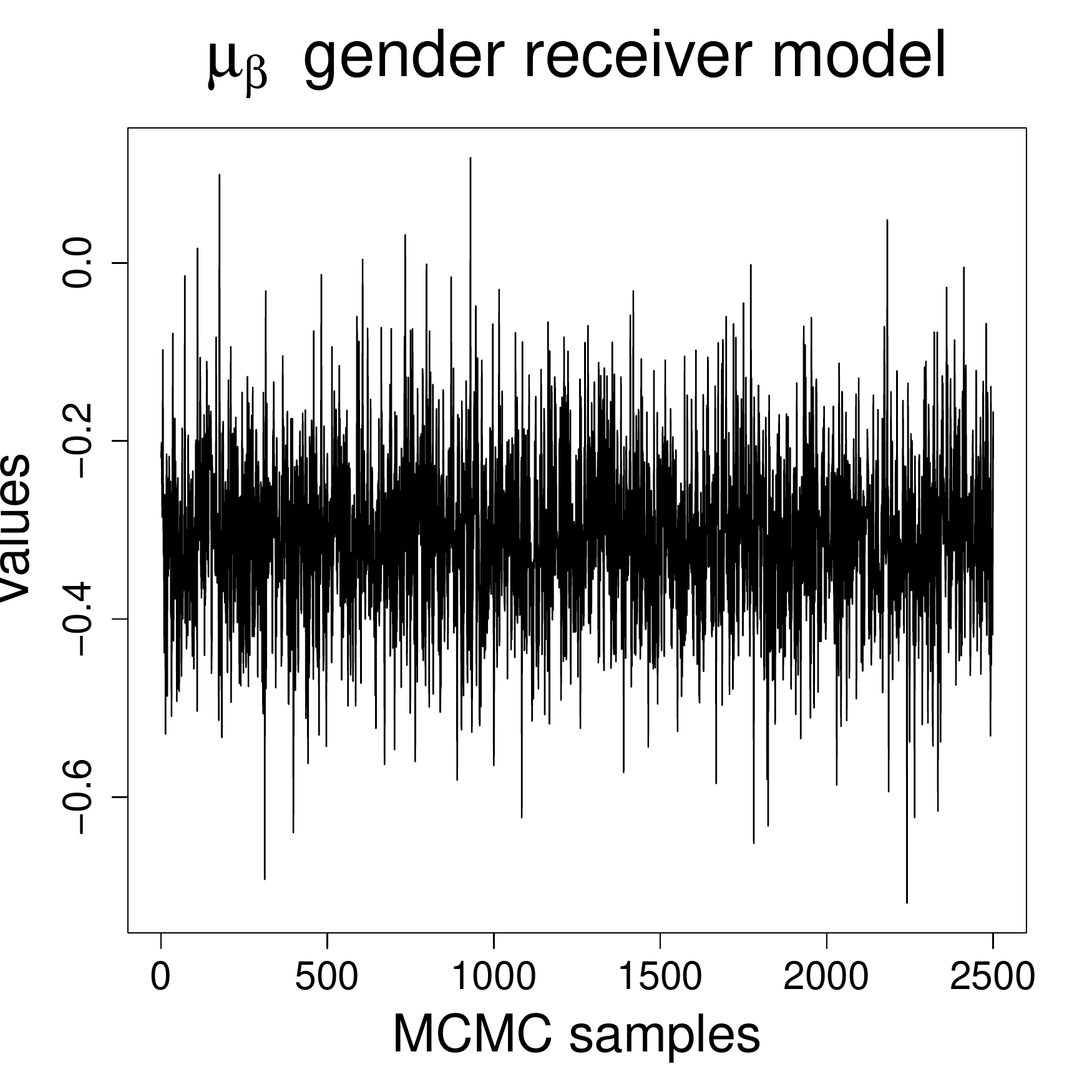}} 
\subfloat{\includegraphics[width = 2in]{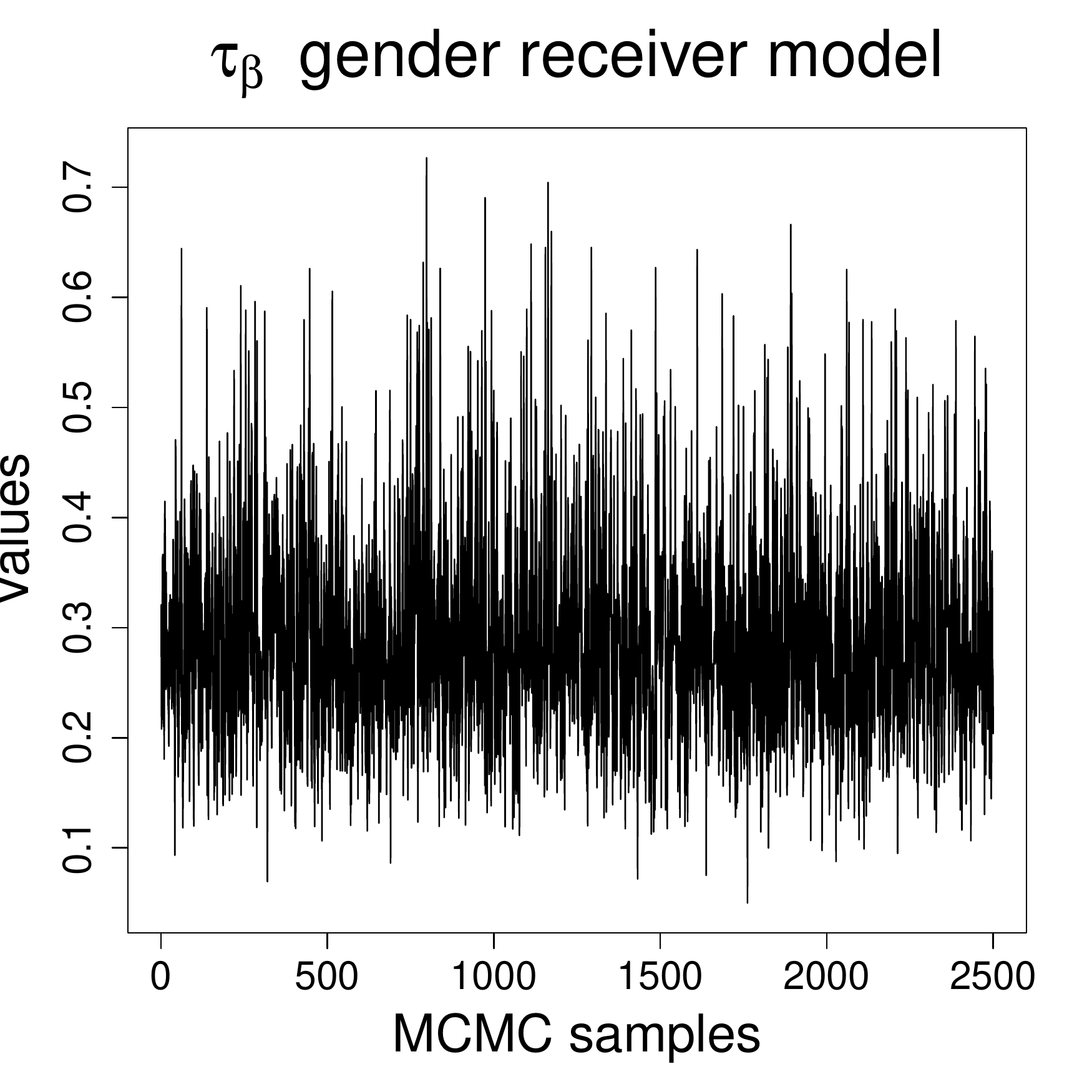}} 

\caption{Trace plots of some of the population parameters in the mixed-effects model. The top (bottom) line show parameters in the sender (receiver) model. The figure display one fixed effect and one random-effect mean with its variance for each model.}
\label{fig:trace_plots_mix}

\end{figure}

In this subsection, we carry out the other proposed tests under the mixed-effects model obtained from the homogeneity tests that were carried out in the previous section. We only restrict ourselves to the results for the full event sequences to keep the discussion of the results as concise as possible. We ran $4$ chains, each with $5000$ MCMC iterations with a burn-in of $2500$ draws. Figure \ref{fig:trace_plots_mix} shows the trace plots of a few parameters, all chains show convergence. Here we also used the $\hat{R}$ as a convergence measure, they were essentially 1 in all chains for all parameters. In addition, the bulk and tail effective sample sizes were all close to around 2000 samples.

\par
\begin{figure}[t]
    \centering
    \subfloat{\includegraphics[width=12cm,height=9cm]{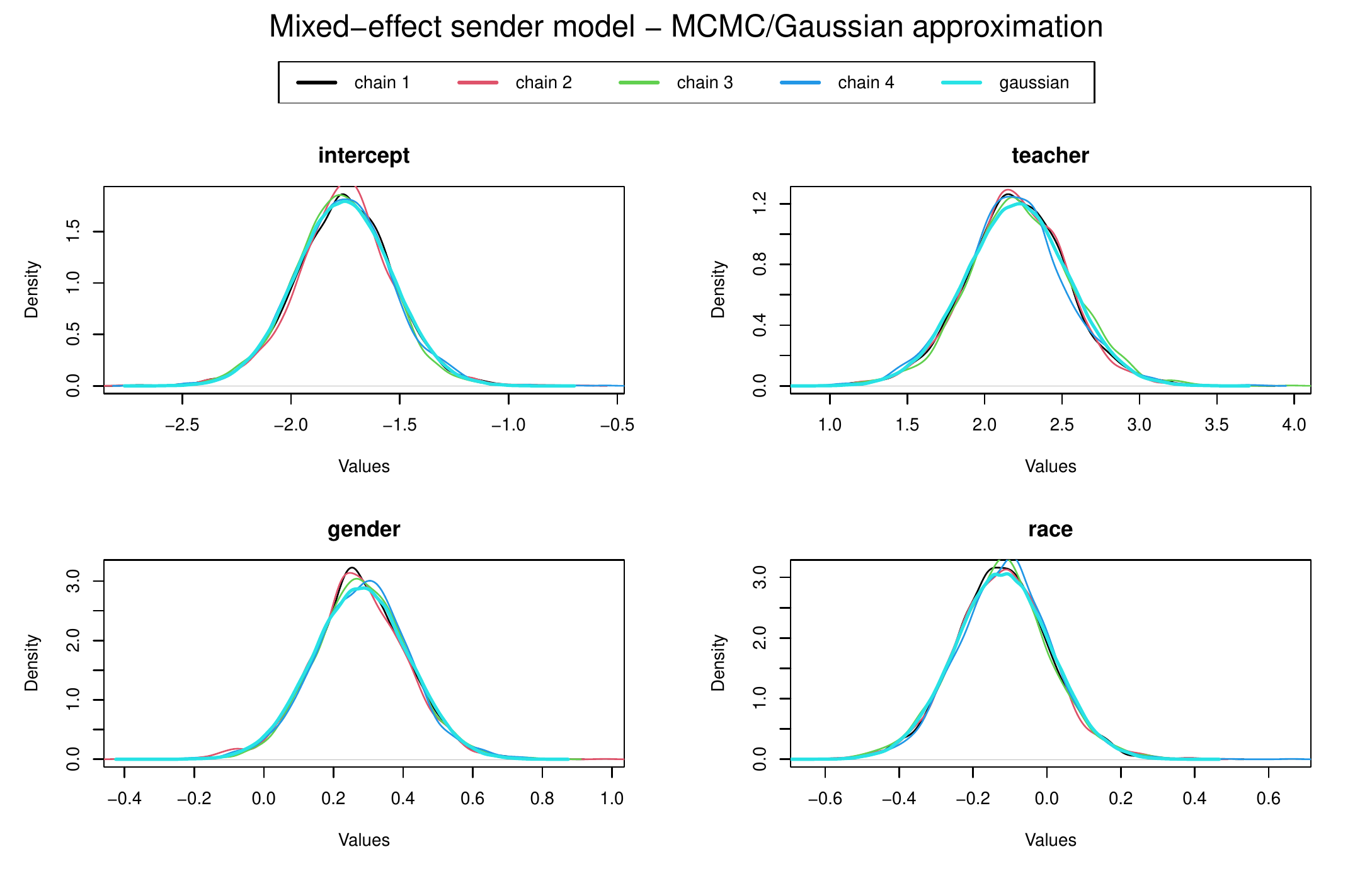}}
    \caption{Estimated posterior density and Gaussian approximation for a some fixed effects in the sender model.}
    \label{fig:mcmc_gauss_app_sender}
\end{figure}
\begin{figure}[t]
    \centering
    \subfloat{\includegraphics[width=12cm,height=9cm]{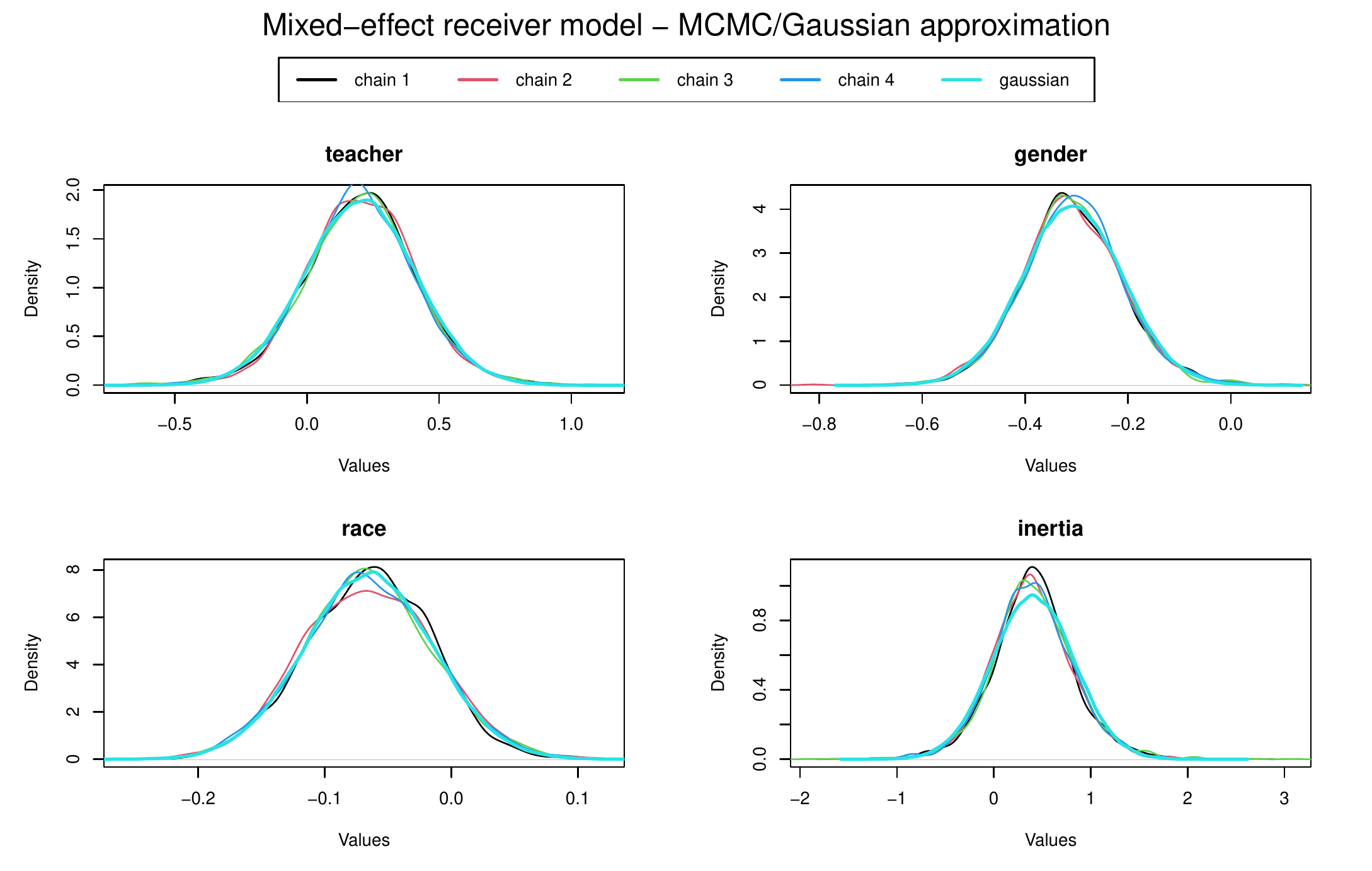}}
    \caption{Estimated posterior density and Gaussian approximation for a some fixed effects in the receiver model.}
    \label{fig:mcmc_gauss_app_receiver}
\end{figure}

Moreover, Figures \ref{fig:mcmc_gauss_app_sender} and \ref{fig:mcmc_gauss_app_receiver} display the density estimated from each of the 4 chains and a Gaussian approximation on top. As can be seen, the Gaussian approximation seems to be acceptable in the bulk as well as in the tails of the distribution. Therefore, empirically, the approach of conducting the tests using these approximations as proxies for the posterior distributions seems reasonable. From now on, given that the results for the four chains are virtually identical, we proceed to analyze the results of a single chain.

\begin{figure}[t]
    \centering
    \includegraphics[width=8cm,height=7cm]{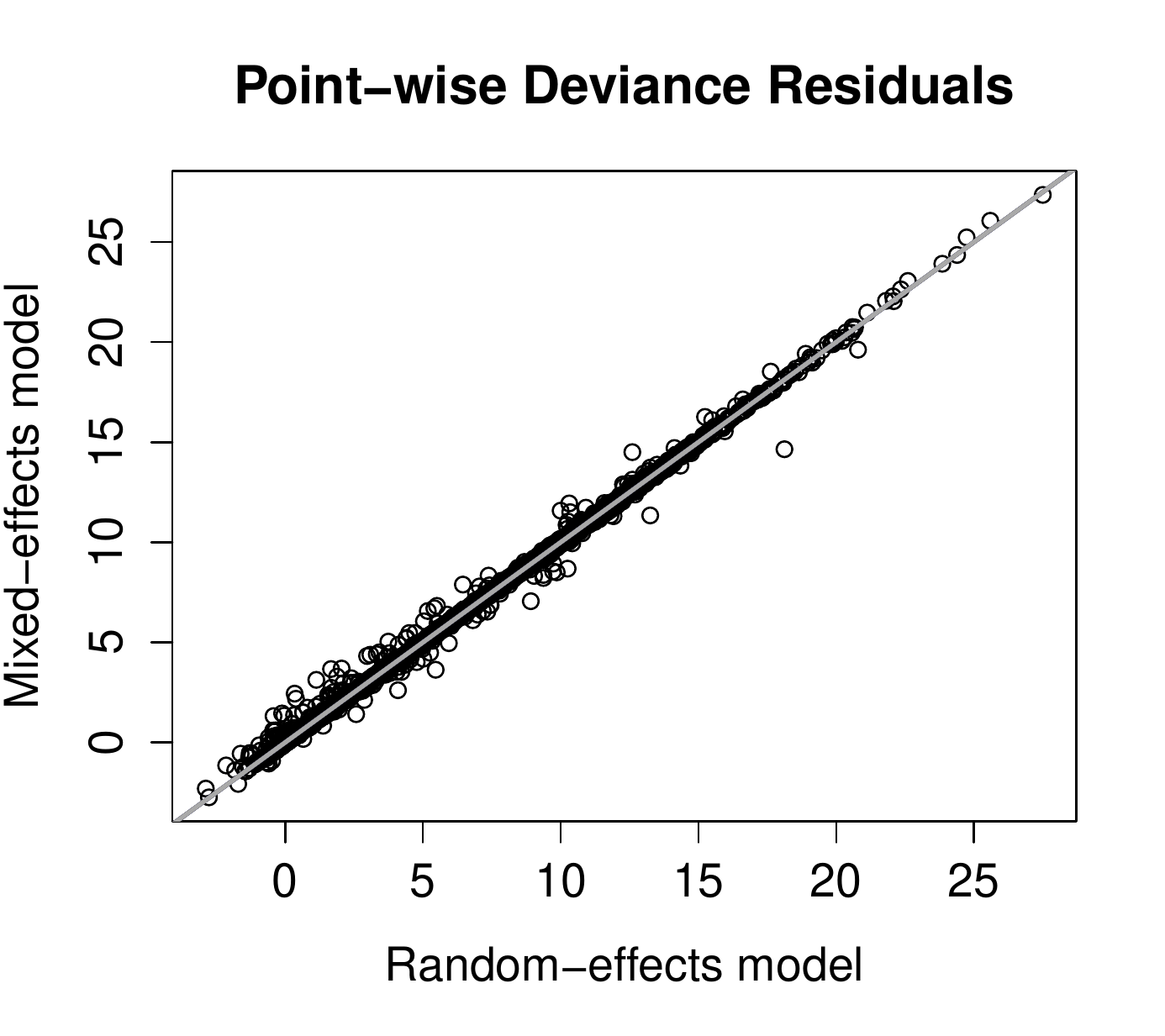}
    \caption{Comparison of the point-wise deviance residuals between the random-effects (x-axis) and the mixed-effects (y-axis) models. The grey line is the equality line.}
    \label{fig:pwdeviance}
\end{figure}

We can see that the proposed testing procedure to determine which effects are random and which are fixed works correctly because a similar fit to the data is obtained in the more parsimonious model (where some effects are fixed) compared to the larger (i.e. all random) model. We can see this by computing the point-wise deviance residuals for both models under the full multilevel model where all effects are random across classrooms and the mixed effects model where certain effects are assumed fixed across classrooms (according to the proposed Bayes factor test). If a similar fit is obtained for both models, the point-wise deviance residuals for both models are similar meaning that none of the two models dominates the other in terms of fit. Figure \ref{fig:pwdeviance} clearly shows that that is the case: most points lie around the grey diagonal where the values are equal. This indicates that the fit provided by both models is similar. Details of the point-wise deviance computation are provided in appendix \ref{app:D}. Finally, table \ref{tab:tab1} shows a comparison of the posterior estimates for both models, where the bold ones are effects treated as fixed in the mixed-effects model. In sum, the mixed effects model results in a comparable fit but with much less parameters to estimate, resulting in a more parsimonious model and a simpler explanation of social interaction behavior across classrooms.

\begin{landscape}
\begin{table}[H]
\centering
\begin{tabular}{ l c c c c }  \hline \hline
 \multicolumn{5}{c}{Parameter Estimates of Mean Effects} \\
 \hline
 & \multicolumn{2}{c}{Random-Effects Model} &  \multicolumn{2}{c}{Mixed-Effects Model} \\
 \cline{2-5}
 Effect & Sender & Receiver & Sender & Receiver\\
 \hline
 intercept & -1.75\ (-2.08, -1.37) & - & -1.77 (-2.19, -1.32) & - \\
 teacher &   2.22\ (1.56, 2.83) & 0.21\ (-0.25, 0.67) & 2.22 (1.60, 2.97) & 0.20 (-0.21, 0.65)\\
 gender & 0.27\ (0.01, 0.53) & -0.30\ (-0.51, -0.12) & 0.29 (0.02, 0.57) & -0.31\ (-0.49, -0.09) \\
 race  & -0.12 (-0.36, 0.11) & -0.07\ (-0.18, 0.05) & -0.12 (-0.37, 0.14) &  \bf{-0.06 (-0.17, 0.04)}\\
 inertia &  -   & 0.42\ (-0.38, 1.15) & - & 0.36 (-0.46, 1.17) \\
 ABB (ABBA) &   1.06\ (0.67, 1.46) & 4.30 (3.88, 4.72) & 1.08 (0.67, 1.52) & 4.27 (3.79, 4.70)\\
 ABA (ABAB) & -0.06\ (-0.52, 0.33)  & 1.29\ (0.57, 1.84) & \bf{-0.05\ (-0.42, 0.32)} & 1.18 (0.53, 1.81)\\
 outgoingness & 0.05 (-0.08, 0.20)  & - & \bf{0.07\ (0.00, 0.12)} & - \\
 popularity & 0.25\ (0.12, 0.39)  & 0.58\ (0.40, 0.77) & \bf{0.17\ (0.11, 0.23)} & 0.59 (0.39, 0.78)\\
 \hline \hline
 
\end{tabular}

\caption{Parameter estimates of the mean effects. Numbers in bold are effects that were treated as fixed, either $\bm{\phi}$ or $\bm{\psi}$, in the mixed-effects model. Not-bolded numbers represent random-effect means, $\bm{\mu}$. Numbers between parentheses are 95\% credible intervals. Dashes indicate that the covariate was not included in the respective model.}
\label{tab:tab1}

\end{table}

\end{landscape}

\subsubsection{Testing the degree of heterogeneity of network effects across school classes}

\par
By testing order constraints on the random effects variance parameters we can assess which characteristics cause most variation in the social interaction behavior across school classes. In particular, we focus on the random effects of the \textbf{\textit{teacher}}, \textbf{\textit{gender}}, and \textbf{\textit{race}} variables in the sender model (which are all dummy variables). The same test can also be applied to the variance parameters in the receiver model in a similar manner. The objective is to determine the amount of evidence regarding the level of heterogeneity among these effects. Let $\sigma^2_{{teacher}}$, $\sigma^2_{{gender}}$, and $\sigma^2_{{race}}$ be the between classrooms variances of the \textbf{\textit{teacher}}, \textbf{\textit{gender}}, and \textbf{\textit{race}} effects, respectively, where these parameters are extracted from the diagonal of $\bm{\Sigma}_{\gamma}$. Given the special (dominant) role of the teacher in a classroom, and the fact that different teachers have different teaching styles, it is expected that the variance of the teacher effect is largest. We tested all six possible order hypotheses for the random effects variances:

\begin{equation*}
\begin{split}
    \text{H}_{1}&: \sigma^2_{teacher} > \sigma^2_{gender} > \sigma^2_{race}\\
    \text{H}_{2}&: \sigma^2_{teacher} > \sigma^2_{race} > \sigma^2_{gender}\\
    \text{H}_{3}&: \sigma^2_{race} > \sigma^2_{teacher} > \sigma^2_{gender}\\
    \text{H}_{4}&: \sigma^2_{race} > \sigma^2_{gender} > \sigma^2_{teacher}\\
    \text{H}_{5}&: \sigma^2_{gender} > \sigma^2_{teacher} > \sigma^2_{race}\\  
    \text{H}_{6}&: \sigma^2_{gender} > \sigma^2_{race} > \sigma^2_{teacher}.\\ 
\end{split}
\end{equation*}

\noindent
The goal is to determine which hypothesis receives most evidence from the data. Below the Bayes factors for each one of these hypotheses against one another is represented in the form of an evidence matrix:

\begin{center}
$\begin{pNiceMatrix}[first-row,first-col]
    & \text{H}_1 & \text{H}_2 & \text{H}_3 & \text{H}_4 & \text{H}_5 & \text{H}_6   \\
\text{H}_1 & 1.00 & 1.47 & 5788.40 & 615.79 & 5788.40 & 275.64 \\
\text{H}_2 & 0.73 & 1.00 & 4200.00 & 446.81 & 4200.00 & 200.00 \\
\text{H}_3 & 0.00 & 0.00 & 1.00 & 0.00 & 0.00 & 0.00\\
\text{H}_4 & 0.00 & 0.00 & 8.40 & 1.00 & 8.40 & 0.40\\
\text{H}_5 & 0.00 & 0.00 & 0.00 & 0.00 & 1.00 & 0.00 \\
\text{H}_6 & 0.00 & 0.00 & 20.00 & 2.13 & 20.00 & 1.0
\end{pNiceMatrix}$
\end{center}
Each cell of the matrix represent the comparison of the hypothesis in the rows against the hypothesis in the columns. So, for example, the evidence for $\text{H}_1$ against $\text{H}_2$ is equal to $1.5$. The evidence for $\text{H}_2$ against $\text{H}_1$ is then $1/1.50 = 0.67$.

\par
The evidence matrix clearly shows that the evidence for hypotheses $\text{H}_1$ and $\text{H}_2$ is much larger than the evidence for any of the other hypotheses, which indicates that the \textbf{\textit{teacher}} effect shows indeed most heterogeneity across the fifteen classrooms. When inspecting the estimates of the variances, this is also confirmed: $\bar{\sigma}^2_{teacher} = 1.65$, $\bar{\sigma}^2_{gender} = 0.25$, and $\bar{\sigma}^2_{race} = 0.22$. Note that the added value of the Bayes factor complementary to eyeballing the estimates is that the effect sizes and their uncertainty are combined in a principled manner to determine which effects are most heterogeneous across classrooms. Finally note that there is no clear evidence that either the gender effect or the race effect is more heterogeneous across classrooms. Thus, we can conclude that teachers have the biggest impact on the variability of social interaction behavior across classrooms.

\subsubsection{Testing the impact of nodal characteristics on classroom dynamics}

\par
In this subsection, we test the effects of nodal characteristics how they affect classroom dynamics. 
In the sender model, due to the special role of the teacher to control interaction behavior during lectures, the \textbf{\textit{teacher}} variable is expected to have the strongest impact on the rate at which an actor starts an interaction when compared to the other two personal-trait variables \textbf{\textit{gender}} and \textbf{\textit{race}}. No expectations are formulated about which of these latter two variables has the largest effect in the model, or about which category of these dichotomous variables results in a positive effect. This results in the following hypotheses where constraints are formulated on the absolute values of the respective effects of these variables

\begin{equation*}
\begin{split}
    \text{H}_{1}&: |\zeta_{{teacher}}| > |\zeta_{{gender}}| > |\zeta_{{race}}|\\
    \text{H}_{2}&: |\zeta_{{teacher}}| > |\zeta_{{gender}}| = |\zeta_{{race}}|\\
    \text{H}_{3}&: |\zeta_{{teacher}}| > |\zeta_{{race}}| > |\zeta_{{gender}}|\\
    \text{H}_{4}&: \text{neither $H_1$, $H_2$, nor $H_3$},
\end{split}
\end{equation*}
The complement hypothesis $H_4$ is included as a ``safety net'' in case our expectations about the dominance of the teacher variable would not be supported by the data.
Once again, we present the resulting fractional Bayes factors in an evidence matrix form
\begin{center}
$\begin{pNiceMatrix}[first-row,first-col]
    & \text{H}_1 & \text{H}_2 & \text{H}_3 & \text{H}_4    \\
\text{H}_1 &  1.00 & 0.21  &   4.74   &  2.39\text{e}8  \\
\text{H}_2 &  4.56 & 1.00  &   21.63  &  1.09\text{e}9  \\
\text{H}_3 &  0.21 & 0.05  &  1.00   & 5.05\text{e}7  \\
\text{H}_4 &  0.00 & 0.00 & 0.00 & 1.00
\end{pNiceMatrix}$.
\end{center}
Again, we translate these Bayes factors to posterior probabilities when assuming that each hypothesis is equally likely a priori, which results in posterior probabilities of 0.235, 0.723, 0.042, and 0.000, for $H_1$, $H_2$, $H_3$, and $H_4$, respectively.
Based on these results we can safely rule out the complement hypothesis $H_4$. Moreover, hypothesis $H_2$ (which assumes that the teacher variable has the largest effect, and the gender effect and the race effect are equal) receives most evidence; but only approximately 3 times more evidence than the hypothesis $H_1$ (which assumes that gender plays a larger role than race after the teacher effect). In order to draw more decisive conclusions of whether $H_1$ or $H_2$ is true more data would be required. These results are confirmed when looking at the posterior estimates, i.e., 2.22, 0.289, 0.143, and their posterior standard deviations, i.e., 0.344, 0.131, 0.103, of the absolute values of the teacher, gender, and race effect. Note again that the Bayes factors and posterior probabilities are a principled probabilistic methodology to summarize these findings without requiring subjective eyeballing of the estimates.


\par
Under the receiver model, we consider a more exploratory approach by considering all possible order hypotheses on the absolute effects of these nodal characteristics. Note that even though the teacher still has a dominant role, this may not necessarily be the case in the receiver model. The following hypotheses are considered

\begin{equation*}
\begin{split}
    \text{H}_{1}&: |\mu_{{teacher}}| > |\mu_{{gender}}| > |\mu_{{race}}|\\
    \text{H}_{2}&: |\mu_{{teacher}}| > |\mu_{{race}}| > 
    |\mu_{{gender}}|\\
    \text{H}_{3}&: |\mu_{{race}}| > |\mu_{{teacher}}| > 
    |\mu_{{gender}}|\\
    \text{H}_{4}&: |\mu_{{race}}| > |\mu_{{gender}}| > 
    |\mu_{{teacher}}|\\
    \text{H}_{5}&: |\mu_{{gender}}| > |\mu_{{race}}| > 
    |\mu_{{teacher}}|\\
    \text{H}_{6}&: |\mu_{{gender}}| > |\mu_{{teacher}}| > 
    |\mu_{{race}}|\\
\end{split}
\end{equation*}

\noindent
where $|\mu|$ is the absolute value of $\mu$, which represents the global mean of the random effects in the receiver model. We also display the results in the form of an evidence matrix which yields:

\begin{center}
$\begin{pNiceMatrix}[first-row,first-col]
   &   H_1    &  H_2   &    H_3   &   H_4  &  H_5  &  H_6\\
H_1 &1.00  & 45.52 & 354.37 & 245.63 & 3.82 & 0.44\\
H_2 &0.02  & 1.00  &  7.78 &  5.39 & 0.08 & 0.01\\
H_3 &0.00  & 0.12  & 1.00 &  0.69 & 0.01 & 0.00\\
H_4 &0.00  & 0.18  &  1.44 &  1.00 & 0.02 & 0.00\\
H_5 &0.26  & 11.91 & 92.69      & 64.25 &1.00 & 0.12\\
H_6 &2.28  & 103.89    & 888.80     & 560.61     & 8.73 &1.00
\end{pNiceMatrix}$.
\end{center}
and translating these to posterior probabilities result in the following posterior probabilities for the respective order hypotheses:  0.291, 0.006, 0.001, 0.001, 0.068, 0.633. These results indicate that almost all probabilities mass goes to $H_1$ and $H_6$, which suggests that the race variable has the smallest impact. Furthermore, the gender variable is expected to play the largest role in the receiver model among these nodal characteristics. These results are confirmed when looking at the posterior estimates, which equal
0.243, 0.310, 0.070, and posterior standard deviations
0.172, 0.100, and 0.045, for the absolute teacher, gender, and race effect, respective.

%
%

%

\subsubsection{Testing the impact of network statistics on classroom dynamics}
\par
Next we focus on testing the effects of endogenous (network) characteristics of actors. Specifically, we evaluate whether the fixed effects of popularity or outgoingness makes actors more likely to be the next sender. 
Let $\phi_{\text{popularity}}$ be the mean effect of \textbf{\textit{popularity}} and $\phi_{\text{outgoingness}}$ be the mean \textbf{\textit{outgoingness}} effect, both in the sender model. We test the following hypotheses against one another:
\begin{equation*}
\begin{split}
    \text{H}_{1}&: \phi_{\text{popularity}} = \phi_{\text{outgoingness}}\\
    \text{H}_{2}&: \phi_{\text{popularity}} < \phi_{\text{outgoingness}}\\
    \text{H}_{3}&: \phi_{\text{popularity}} > \phi_{\text{outgoingness}}.
\end{split}
\end{equation*}

\begin{figure}[t]
    \centering
    \includegraphics[width=8cm,height=6cm]{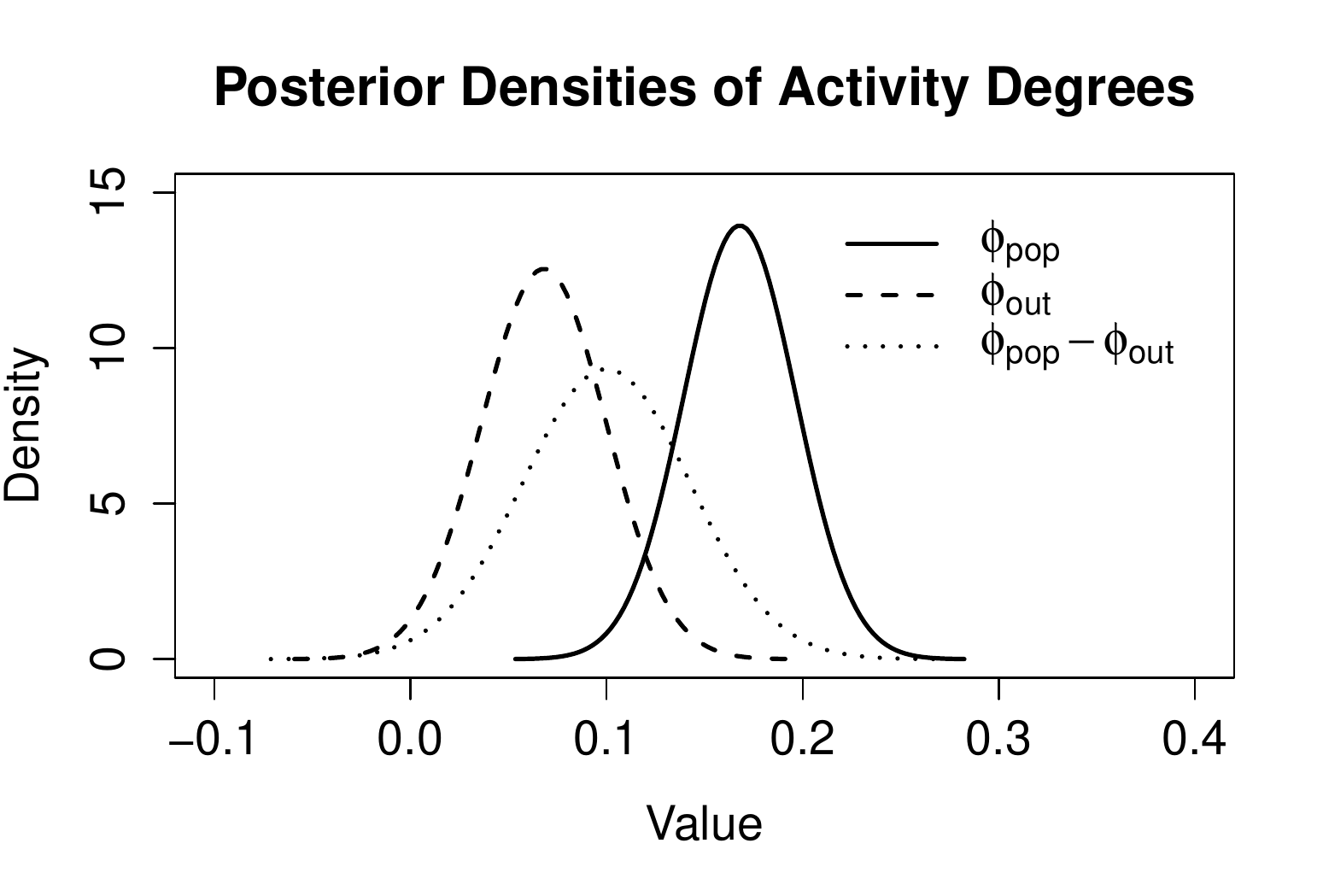}
    \caption{Posterior distribution of $\phi_{\text{popularity}}$ (solid line) and $\phi_{\text{outgoingness}}$ (traced line) and their difference (dotted line).}
    \label{fig:all_hyp}
\end{figure}

\noindent
Having more than two hypotheses, we show the Bayes factors in an evidence matrix form, which yields
\begin{center}
$\begin{pNiceMatrix}[first-row,first-col]
    & \text{H}_1 & \text{H}_2 & \text{H}_3    \\
\text{H}_1 &  1.00  &   48.55   &  0.43  \\
\text{H}_2 &  0.02  &  1.00  &   0.01  \\
\text{H}_3 &  2.29  &  110.99   & 1.00   
\end{pNiceMatrix}$.
\end{center}
Assuming equal prior probabilities of the three hypotheses, the posterior probabilities are equal to $P(\text{H}_1|\textbf{E})=0.384$, $P(\text{H}_2|\textbf{E})=0.006$, and $P(\text{H}_3|\textbf{E})=0.610$. The results indicate that $\text{H}_{3}$ is most likely to be true after observing the data, being approximately twice more likely than $\text{H}_{1}$, and approximately 100 times more likely than $\text{H}_2$. For this reason, we can with almost complete certainty state that the popularity effect is not smaller than the outgoingness effect. This is confirmed by the posterior estimates of the popularity and the outgoingness effect which are equal to 0.17 and 0.07, respectively; see also the posteriors in Figure \ref{fig:all_hyp}. More data is required in order to obtain more decisive evidence of whether $\text{H}_{1}$ or $\text{H}_{3}$ is true. 

\par
In addition to testing the fixed effects as above, we can also test the global means of random effects. Here, we test the participation shift \textbf{ABBA} is more likely than the participation shift \textbf{ABAB} (both are random-effects) after controlling for the other effects in the receiver model. 
Let $\mu_{\text{abba}}$ be the global mean effect of \textbf{ABBA} and $\mu_{\text{abab}}$ be the global mean effect of \textbf{ABAB}. This comparison captures the tension between an individual's tendency to continue to speak to the same individual or the societal norm of reciprocity where the previous receiver because the sender of the next event and the previous sender because the receiver. In polite conversation norms, we would expect to see more turn-switches between two individuals (A speaks to B and then B responds to A) than turn-continuing (where A directs a comment to B and then continues to speak without giving B the opportunity to respond first). 
The estimates in Table \ref{tab:tab1} suggests that this is indeed the case since the mean effect of \textbf{ABBA} is almost four times as large as the mean effect of \textbf{ABAB}, 4.27 and 1.18, respectively. Here we formally test this by considering the following three hypotheses:
\begin{equation*}
\begin{split}
    \text{H}_{1}&: \mu_{\text{abba}} = \mu_{\text{abab}}\\
    \text{H}_{2}&: \mu_{\text{abba}} < \mu_{\text{abab}}\\
    \text{H}_{3}&: \mu_{\text{abba}} > \mu_{\text{abab}}.
\end{split}
\end{equation*}

This yields the following evidence matrix:
\begin{center}
$\begin{pNiceMatrix}[first-row,first-col]
    & \text{H}_1 & \text{H}_2 & \text{H}_3    \\
\text{H}_1 &  1.00  &   143.26   &  0.00  \\
\text{H}_2 &  0.00  &  1.00  &   0.00  \\
\text{H}_3 &  4.19 \times 10^{12}  &  6.00 \times 10^{14}   & 1.00   
\end{pNiceMatrix}$.
\end{center}
The evidence favoring $\text{H}_{3}$ is overwhelmingly larger than the evidence for any other hypothesis, strongly supporting the hypothesis that the effect of immediate reciprocity is larger than that of immediate inertia across these fifteen classrooms at Magnet High school. This is also confirmed by the posterior probabilities $P(\text{H}_1|\textbf{E})=0.000$, $P(\text{H}_2|\textbf{E})=0.000$, and $P(\text{H}_3|\textbf{E})=1.000$, when assuming equal prior probabilities.

\par
Finally we showcase a hypothesis test of network effects between the receiver model and the sender model. Specifically, we test whether the \textbf{\textit{teacher}} variable has a larger effect on the rate of being a sender or on the rate of being chosen as receiver. Given the dominant role of the teacher as a sender, we expect the teacher effect is larger in the sender model. 
We formulate the hypotheses as follows:

\begin{equation*}
\begin{split}
    \text{H}_{1}&: \zeta_{{teacher}} = \mu_{{teacher}}\\
    \text{H}_{2}&: \zeta_{{teacher}} < \mu_{{teacher}}\\
    \text{H}_{3}&: \zeta_{{teacher}} > \mu_{{teacher}}.
\end{split}
\end{equation*}
where hypothesis $H_3$ corresponds to our scientific expectation.

\noindent
The resulting evidence matrix for the Bayes factors is: 

\begin{center}
$\begin{pNiceMatrix}[first-row,first-col]
    & \text{H}_1 & \text{H}_2 & \text{H}_3    \\
\text{H}_1 &  1.00  &   92.15   &  0.00  \\
\text{H}_2 &  0.01  &  1.00  &   0.00  \\
\text{H}_3 &  5.84 \times 10^{4}  &  5.38 \times 10^{6}   & 1.00   
\end{pNiceMatrix}$.
\end{center}

\noindent
Indeed, the evidence matrix shows convincing evidence that the teacher effect is larger in the sender model than in the receiver model. The posterior probabilities of the three hypotheses are equal to $P(\text{H}_1|\textbf{E})=0.000$, $P(\text{H}_2|\textbf{E})=0.000$, and $P(\text{H}_3|\textbf{E})=1.000$, when assuming equal prior probabilities, which results in the same conclusion. The results are also confirmed by the estimates which are equal to $\bar{\zeta}_{{teacher}} = 2.22$ and $\bar{\mu}_{{teacher}} = 0.20$.

\section{Discussion}\label{sec:disc}

\par
In this paper we presented a Bayesian actor-oriented multilevel relational event model for studying social interaction behavior from independent relational event sequences. This model allows for inferences at the actor level, thus opening the possibility of unveiling effects that make an actor more prone to send or receive an interaction in the population under study. Our results show that the model is able to capture the effects that have the largest impacts in actors preferences, even when those effects are different in the sender activity and receiver choice rates. The models can be estimated using the Stan programming language, the Stan code for the actor-oriented relational event model is available on Github (\textcolor{red}{link suppressed for peer review}). 

\par
Furthermore, a flexible set of hypothesis testing procedures was proposed for this class of models, facilitating inferences on population parameters of the estimated effects. These tests can be used for testing whether effects should be treated as fixed or random across sequences, for testing the relative heterogeneity of different network effects across sequences, and for testing equality and order constraints on the fixed effects, even on their absolute values. These tests are useful to get a better understanding about the heterogeneity of social interaction behavior across independent relational event sequences. These novel testing procedures can also be applied to other mixed effects models to better understand the heterogeneity in other types of multilevel or clustered data. 

\par
The computational issues originating from the inefficiencies that result from the hierarchical structure of the model are the main limitations in the estimation process. We have taken advantage of the multivariate normal structure of the hierarchical prior to induce a non-centered transformation in the random-effects parameters, which is more efficient in practice for the reasons aforementioned. An issue that we have not directly addressed in our paper is the sparsity of relational event data which may cause  the geometry of the posterior distribution to be highly complex (\cite{betancourt2015hamiltonian}). The alien form of the likelihood of the relational event model also presents a challenge to the estimation of hierarchical relational event models. An especially attractive next step would to construct alternative representations of the model so that posterior distributions in closed form could be derived, improving efficiency in the estimation process.

Another promising direction to improve computational feasibility when fitting multilevel relational event models to the large clustered relational event sequences is using meta-analytic approximations \citep{borenstein2010basic}.
A Gibbs sampler could be derived using the point estimates from the independent relational event sequences as data observations. The question would then be how large the relational event sequences should be in order for these meta-analytic approximation to be accurate enough to make reliable statistical inferences. Our flexible Bayes factor tests could then be built on top of that. Finally, another important direction for future research would be to model time-varying coefficients across sequences in order to discover complex temporal changes of the network drivers of social interactions over time.

\clearpage

\appendix

\section{Actor-oriented relational event model as a Poisson regression model}\label{app:AO-Poisson}

\par
The likelihood of the actor-oriented relational event model is a product of a piece-wise constant exponential likelihood and a multinomial likelihood. Where the former corresponds to the sender model and the latter to the receiver model. Both of these models are special cases of the Poisson model.

\paragraph{Sender model as a Poisson regression:}

\begin{proof}
Assuming $M$ events are observed, and the risk set contains $N$ actors, 
\begin{align*}
    P (\textbf{E}\ |\ \bm{\gamma}  ) &= \prod_{m = 1}^{M} P \Big((s_m, t_m) | \bm{\gamma}, (e_{1}, \dots, e_{m-1}) \Big)  = \\
    &= \prod_{m = 1}^{M} \lambda_{s_m}\ \times \exp \{-(t_{m} - t_{m-1}) \sum_{s \in \mathcal{R}} \lambda_{s}\} \\
    &= \prod_{m = 1}^{M} \exp \{\bm{\gamma} \bm{x}'_{s_m} \} \times \exp \{ - \exp \{\upsilon_{m} \} \sum_{s \in \mathcal{R}} \exp \{\bm{\gamma} \bm{x}'_{s_m} \}  \} &&\big(\upsilon_{m} = \log(t_{m} - t_{m-1})\big)\\
    &\propto \prod_{m = 1}^{M} \exp \{\upsilon_{m} + \bm{\gamma} \bm{x}'_{s_m} \} \times \prod_{s \in \mathcal{R}} \exp \{ - \exp \{\upsilon_{m} + \bm{\gamma} \bm{x}'_{s_m} \}  \} \\
    &= \prod_{m = 1}^{M} \prod_{s \in \mathcal{R}} {\exp \{\upsilon_{m} + \bm{\gamma} \bm{x}'_{s_m} \}}^{y_{s}} \times \exp \{ - \exp \{\upsilon_{m} + \bm{\gamma} \bm{x}'_{s_m} \}  \} \\
    &= \prod_{m = 1}^{M} \prod_{s \in \mathcal{R}} P \Big( y_{s_m} | \lambda_{s_m}= \exp \{\upsilon_{m} + \bm{\gamma} \bm{x}'_{s_m } \} \Big),
\end{align*}
where $y_{s_{m}} = 1$ if actor $s$ is observed and zero otherwise. As a result, the factorial term in the Poisson likelihood will always be equal to one, since $1! = 1$.
\end{proof}

\paragraph{Receiver model as a Poisson regression:}

\begin{proof}
Once again, assuming $M$ events are observed, and the risk set contains $N$ actors, the likelihood function of the receiver model has the following form
\begin{align*}
    P(\textbf{E} | \bm{\beta} ) &= \prod_{m=1}^{M} \frac{\lambda_{r_{m} | s_{m}}}{\sum_{r \in \mathcal{R}} \lambda_{r | s_{m}}} = \prod_{m=1}^{M} \frac{\exp \{ \bm{\beta} \bm{x}'_{s_{m} r_{m}} \}}{\sum_{r \in \mathcal{R}} \exp \{ \bm{\beta} \bm{x}'_{s_m r} \}}.
\end{align*}
Taking the logarithm of $p(\textbf{E} | \bm{\beta})$ yields
\begin{align*}
    \log(p(\textbf{E} | \bm{\beta})) = \sum_{m=1}^{M} \bm{\beta} \bm{x}'_{s_{m} r_{m}} - \Big( \sum_{r \in \mathcal{R}} \exp \{ \bm{\beta} \bm{x}'_{s_m r} \}  \Big) \tag{$\star$}
\end{align*}
Then, when considering an alternative formulation using a Poisson model, an equivalent likelihood can be written as
\begin{align*}
    \tilde{\bm{Y}}_{r_{m} | s_{m}} \sim \text{Poisson} \Big( \exp \{ \bm{\beta} \bm{x}'_{s_{m} r_{m}} + \alpha_{m} \} \Big).
\end{align*}
Where $\alpha_{m}$ is an event-specific intercept and $\tilde{Y}_{r | s } = 1$ if the dyad $(s, r)$ was observed and zero otherwise. The likelihood of the Poisson model can then be written
\begin{align*}
    P(\tilde{\bm{Y}}_{r | s} | \bm{\beta}, \bm{\alpha}) &= \prod_{m = 1}^{M} \prod_{r \in \mathcal{R}} {\Big( \exp \{ \bm{\beta} \bm{x}'_{s_{m} r_{m}} + \alpha_{m}\} \Big)}^{\tilde{y}_{r_m | s_m}} \times\\
    &\times \exp \{ - \exp \{ \bm{\beta} \bm{x}'_{s_{m} r} + \alpha_{m}\} \} \times {\tilde{y}_{r_m | s_m}}^{-1}
    \end{align*}
Where $\tilde{y}_{r_m | s_m} = 1$. To write the likelihood as a function of $\bm{\beta}$, we need to derive the maximum likelihood estimate of $\alpha_m$. Take the logarithm and deriving with respect to $\alpha_m$ yields the following MLE expression
\begin{align*}
    \hat{\alpha}_{m} = - \log \Big( \sum_{r \in \mathcal{R}} \exp \{ \bm{\beta} \bm{x}'_{s_m r} \} \Big)
\end{align*}
Plug in the MLE into the likelihood
\begin{align*}
    \log P(\tilde{\bm{Y}}_{r | s} | \bm{\beta}, \hat{\bm{\alpha}}) &= \sum_{m=1}^{M} \sum_{r \in \mathcal{R}}  \Big[ \tilde{y}_{r_m | s_m} \big( \bm{\beta} \bm{x}'_{s_m r_m} + \hat{\alpha}_{m} \big) -  \exp \{ \bm{\beta} \bm{x}'_{s_m r} + \hat{\alpha}_{m}\} \} \Big] \\
    &= \sum_{m=1}^{M} \sum_{r \in \mathcal{R}}  \tilde{y}_{r_m | s_m} \bm{\beta} \bm{x}'_{s_m r_m} - \log \Big( \sum_{r \in \mathcal{R}} \exp \{ \bm{\beta} \bm{x}'_{s_m r} \} \Big) - \\
    &- \exp \{ \bm{\beta} \bm{x}'_{s_m r} - \log \big( \sum_{r \in \mathcal{R}} \exp \{ \bm{\beta} \bm{x}'_{s_m r} \} \big) \} \\
    &= \sum_{m=1}^{M} \bm{\beta} \bm{x}'_{s_m r_m} - \log \Big( \sum_{r \in \mathcal{R}} \exp \{ \bm{\beta} \bm{x}'_{s_m r} \} \Big) - \\
    &- \sum_{r \in \mathcal{R}} \exp \{ \bm{\beta} \bm{x}'_{s_m r} \times \exp \{\log {\big( \sum_{r \in \mathcal{R}} \exp \{ \bm{\beta} \bm{x}'_{s_m r} \} \big)}^{-1} \} \} \\
    &= \sum_{m=1}^{M} \bm{\beta} \bm{x}'_{s_m r_m} - \log \Big( \sum_{r \in \mathcal{R}} \exp \{ \bm{\beta} \bm{x}'_{s_m r} \} \Big) - \\
    &- \cancel{\frac{\sum_{r \in \mathcal{R}} \exp \{ \bm{\beta} \bm{x}'_{s_m r} \}}{\sum_{r \in \mathcal{R}} \exp \{ \bm{\beta} \bm{x}'_{s_m r}\}}}\\ 
    &\propto \sum_{m=1}^{M} \bm{\beta} \bm{x}'_{s_m r_m} - \log \Big( \sum_{r \in \mathcal{R}} \exp \{ \bm{\beta} \bm{x}'_{s_m r} \} \Big)
\end{align*}
Which is equal to $(\star)$.
\end{proof}

\section{Linear transformation of multivariate normal distribution}\label{app:A}

\par
Here we show that the transformation of the random-effects parameters is valid. Let $\bm{X}$ be a p-dimensional random variable, with $\bm{X} \sim \mathcal{N}(\bm{\mu}, \bm{\Sigma})$, where $\bm{\mu} \in {\rm I\!R}^p$ is the mean vector and $\bm{\Sigma}$ is a $p \times p$ covariance matrix. Thus, the moment generating function of $\bm{X}$ is given by

\begin{equation*}
    M_{\bm{X}}(\bm{t}) = \exp \Big\{ \bm{t}' \bm{\mu} + \frac{1}{2} \bm{t}' \bm{\Sigma} \bm{t} \Big\}.
\end{equation*}

\noindent
Now, assuming we can write $\bm{X} = \bm{\mu} + \bm{A} \bm{Z}$, where $\bm{A}$ is a $p \times p$ matrix, with $\bm{\Sigma} = \bm{A}' \bm{A}$, and $\bm{Z} = (Z_1, Z_2, \dots, Z_p)$ is an independent normal random vector, with $Z_{i} \sim \mathcal{N}(0, 1),\ \text{for}\ i = 1, \dots, p$. Therefore, if we can show that $M_{\bm{X}}(\bm{t}) =  M_{(\bm{\mu} + \bm{A} \bm{Z})}(\bm{t})$, then the transformation holds. So, we derive the moment generating function of the transformation as

\begin{proof}
\begin{align*}
    M_{(\bm{\mu} + \bm{A} \bm{Z})}(\bm{t}) &=  \text{E}(e^{\bm{t}' \bm{X}}) = \text{E}(e^{\bm{t}' (\bm{\mu} + \bm{A}\bm{Z})})\\
    &= e^{\bm{t}' \bm{\mu}}\ \text{E}(e^{\bm{l}' \bm{Z}}),\ \text{where}\ \bm{l}' = \bm{t}' \bm{A} \\
    &= e^{\bm{t}' \bm{\mu}}\  \text{E}\Big(e^{\sum_{i = 1}^{p} l_{i} Z_{i}}\Big)\\
    &= e^{\bm{t}' \bm{\mu}}\ \prod_{i = 1}^p \text{E}(e^{l_{i} Z_{i}}) \\
    &= e^{\bm{t}' \bm{\mu}}\ \prod_{i = 1}^p e^{l^2_{i}/2} \\
    &= \exp \Big\{ \bm{t}' \bm{\mu}\ + \sum_{i = 1}^{p} \frac{l^2_{i}}{2} \Big\}\\
    &= \exp \Big\{ \bm{t}' \bm{\mu}\ + \frac{1}{2} \bm{l}' \bm{l} \Big\} = \exp \Big\{ \bm{t}' \bm{\mu}\ + \frac{1}{2} \bm{t}' \bm{\Sigma} \bm{t} \Big\}.
\end{align*}
\end{proof}

\par
Thus $M_{\bm{X}}(\bm{t}) =  M_{(\bm{\mu} + \bm{A} \bm{Z})}(\bm{t})$. Hence, we can transform the random-effects parameters in an MCMC algorithm by following the steps,

\begin{enumerate}
    \item Obtain posterior samples of mean vector $\bm{\mu} \in {\rm I\!R}^p$;
    \item Obtain posterior samples of the $p \times p$ covariance matrix $\bm{\Sigma}$;
    \item Compute the Cholesky factor $\bm{A}$ of $\bm{\Sigma}$;
    \item Sample a p-dimensional vector of independent and identically distributed standard normal variables $\bm{Z}$;
    \item Compute the transformation $\bm{\beta} = \bm{\mu} + \bm{A} \bm{Z}$.
\end{enumerate}

\section{Bayes factor of constrained against unconstrained hypotheses}\label{app:B}

\par 
Let us suppose that we wish to test an informative hypothesis $\text{H}_{i}: \bm{\theta} \in \bm{\Theta}_{i}$, where  $\bm{\Theta}_{i}$ can be seen as a truncation of the space of $\bm{\theta} \in \bm{\Theta}$, with $\bm{\Theta}_{i} \subset \bm{\Theta}$. Thus, if we define the prior in the space $\bm{\Theta}_{i}$ as a truncation of the prior in $\bm{\Theta}$, such as $\pi_{i}(\bm{\theta}) = c^{-1} \pi(\bm{\theta})\ \mathbbm{I} \{\bm{\theta} \in \bm{\Theta}_{i}\}$, where $c$ is a normalizing constant of the form $c = \int_{\bm{\theta} \in \bm{\Theta}_{i}} \pi(\bm{\theta}) d\bm{\theta}$ and $\pi(\bm{\theta})$ is the prior under the unconstrained space. Then, the Bayes factor of $\text{H}_{i}$ against $\text{H}_{u}$ will be given by

\begin{equation*}
\begin{split}
    \text{BF}_{iu} &= \frac{m(\textbf{E} | \text{H}_{i})}{m(\textbf{E} | \text{H}_{u})} = \frac{\int_{\bm{\theta} \in \bm{\Theta}_{i}} \pi_{i}(\bm{\theta}) \pi(\textbf{E} | \bm{\theta}) d\bm{\theta}}{\int \pi(\bm{\theta}) \pi(\textbf{E} | \bm{\theta}) d\bm{\theta}}
    = c^{-1} \int_{\bm{\theta} \in \bm{\Theta}_{i}} \frac{\pi(\bm{\theta}) \pi(\textbf{E} | \bm{\theta}) d\bm{\theta}}{\int \pi(\bm{\theta}) \pi(\textbf{E} | \bm{\theta}) d\bm{\theta}} =\\
    &= c^{-1} \int_{\bm{\theta} \in \bm{\Theta}_{i}} \pi(\bm{\theta} | \textbf{E}) d\bm{\theta} = \frac{\int_{\bm{\theta} \in \bm{\Theta}_{i}} \pi(\bm{\theta} | \textbf{E}) d\bm{\theta}}{\int_{\bm{\theta} \in \bm{\Theta}_{i}} \pi(\bm{\theta}) d\bm{\theta}} = \frac{P(\bm{\theta} \in \bm{\Theta}_{i} | \textbf{E})}{P(\bm{\theta} \in \bm{\Theta}_{i})}.
\end{split}
\end{equation*}

\noindent
Therefore, the Bayes factor of a constrained against an unconstrained hypothesis can be reduced to a ratio of posterior and prior probabilities in the space of the constrained hypothesis. In case of an exact hypothesis, such as $\text{H}: \bm{\theta} = \bm{0}$, the ratio of probabilities becomes the Savage-Dickey density ratio \citep{dickey1971weighted}.

\section{Bayes factor for testing random-effect structures}\label{app:C}

\par
In this appendix we show details on the derivation of the parameters in the Bayes factor for testing random-effect structures. Assuming we have $K$ social networks and each of them has its own $\beta_{k},\ \text{for}\ i = 1, \dots, K$, regression parameter. Where $\beta_{k} \sim \mathcal{N}(\mu, \sigma^2)$. Thus, let $\bar{\mu}$, $\bar{\sigma}^2$, $\bar{\beta}_k$ be posterior estimates for $\mu$, $\sigma^2$ and $\beta_k,\ \forall\ k$. Also, let $\bar{\tau}^2_k$ be the point estimate for the variance of the posterior distribution of $\beta_k$. Then we can write

\begin{equation}
\begin{split}
    \text{posterior:}\ &\beta_{k}| \text{E}_{k} \sim \mathcal{N}(\bar{\beta}_{k}, \bar{\tau}^2_{k})\\
    \text{prior:}\ &\beta_{k} \sim \mathcal{N}(\bar{\mu}, \bar{\sigma}^2).
\end{split}
\end{equation}

\noindent
Assuming we want to test $\text{H}_0: \beta_1 = \dots = \beta_K$ against $\text{H}_1: \text{"at least one is different"}$. Thus we can do so by writing $\xi_{j} = \beta_{j+1} - \beta_{j}$, for $j = 1, \dots, K-1$. Therefore we can approximate the joint distribution of $\bm{\xi} = (\xi_1, \dots, \xi_{K-1})$ by a multivariate normal

\begin{equation*}
\begin{split}
    \text{posterior:}\ &\bm{\xi}| \bm{\text{E}} \sim \mathcal{N}(\bar{\bm{\mu}}_{_\xi},\  \bar{\bm{\Sigma}}_{_\xi})\\
    \text{prior:}\ &\bm{\xi} \sim \mathcal{N}(\bm{0},\ \bar{\bm{\Lambda}}_{_\xi}).
\end{split}
\end{equation*}

\noindent
Thus we can derive the parameters of the distributions of every component $j = 1, \dots, K-1$ of $\bm{\xi}$ as follows

\begin{equation*}
\begin{split}
    \text{E}(\xi_{j} | \textbf{E}) &= \text{E}(\beta_{j+1} - \beta_{j})\\
    &= \text{E}(\beta_{j+1}) - \text{E}(\beta_{j}) = \bar{\beta}_{j+1} - \bar{\beta}_{j},\\
\end{split}
\end{equation*}

\begin{equation*}
\begin{split}
    \text{Var}(\xi_{j}| \textbf{E}) &= \text{Var}(\beta_{j+1} - \beta_{j})\\
    &= \text{Var}(\beta_{j+1}) + \text{Var}(\beta_{j}) = \bar{\tau}^2_{j+1} + \bar{\tau}^2_{j},\\
\end{split}
\end{equation*}

\begin{equation*}
\begin{split}
    \text{Cov}(\xi_{j}, \xi_{j+h} | \textbf{E}) & = \text{E}(\xi_{j} \xi_{j+h}) - \text{E}(\xi_{j}) \text{E}(\xi_{j+h})\\
    &= \text{E}\Big((\beta_{j+1} - \beta_{j}) (\beta_{j+h+1} - \beta_{j+h})\Big) - \text{E}\Big((\beta_{j+1} - \beta_{j})\Big) \text{E}\Big((\beta_{j+h+1} - \beta_{j+h})\Big)\\
    &= \text{E}(\beta_{j+1} \beta_{j+h+1}) - \text{E}(\beta_{j+1} \beta_{j+h}) - \text{E}(\beta_{j} \beta_{j+h+1}) + \text{E}(\beta_{j} \beta_{j+h}) - \\
    &- \text{E}(\beta_{j+1}) \text{E}(\beta_{j+h+1}) + \text{E}(\beta_{j+1}) \text{E}(\beta_{j+h}) + \text{E}(\beta_{j}) \text{E}(\beta_{j+h+1}) - \text{E}(\beta_{j}) \text{E}(\beta_{j+h})\\
    &= \begin{cases}
    - \text{Var}(\beta_{j+1}) = - \bar{\tau}^2_{j+1},& \text{if } h = 1\\
    \ \ 0,              & \text{if } h > 1.
\end{cases}
\end{split}
\end{equation*}

\noindent
The parameters in the prior are derived in the same way with $\text{E}(\xi_{j}) = 0$, $ \text{Var}(\xi_{j}) = 2 \bar{\sigma}^2$ and $\text{Cov}(\xi_{j}, \xi_{j+h}) = - \bar{\sigma}^2$, if $h = 1$, and zero otherwise.

\section{Point-wise deviance residuals}\label{app:D}

After estimation, it is common practice to check how the proposed model fits the data under study. In \cite{dubois2013hierarchical}, they used the posterior draws to evaluate the log likelihood at every data point to determine the dynamic adequacy of the model to every data point. Thus, for the actor-oriented relational event model, for $k = 1, \dots, K$ and $m = 1, \dots, M_{k}$, this quantity would be computed as

\begin{equation}\label{dev_actor}
\begin{split}
    \text{Res}_m = -2 &\Bigg\{ \log\Big(\lambda_{s_m}\Big(t_m| \textbf{E}_{k}\Big)\Big)  +  \log\Big(\lambda_{r_m|s_m}\Big(t_m | s_m, \textbf{E}_{k}\Big)\Big) -  \\
    & - \log \Bigg(\sum_{s,r} \lambda_{r|s_m}\Big(t_m| s_m, \textbf{E}_{k}\Big)\Bigg) - (t_m - t_{m - 1}) \Big(\sum_{i} \lambda_{s}(t_m| \textbf{E}_{k}\Big)\Big) \Bigg\}.
\end{split}
\end{equation}

\par
This measure is usually called residual deviance \citep{collett2015modelling}, and it is useful to compare models and see which one fits better each data point by having smaller values of $\text{Res}_m$. 



\bibliography{main}
\bibliographystyle{apacite}


\clearpage

\end{document}